\documentclass[11pt,a4paper,pdflatex]{article}
\usepackage{jheppub}

\usepackage{physics,bm}
\usepackage{amsmath,amsfonts,amssymb}

\usepackage{comment,here}

\usepackage{natbib}

\title{Dynamical stability and filamentary instability in holographic conductors}

\author[a,b]{Shuta~Ishigaki}
\author[b,c]{Shunichiro~Kinoshita}
\author[d]{Masataka~Matsumoto}
\affiliation[a]{Department of Physics, Shanghai University,\\
99 Shangda road, Shanghai 200444, China}
\affiliation[b]{Department of Physics, Chuo University,\\
Kasuga, Bunkyo-ku, Tokyo 112-8551, Japan}
\affiliation[c]{Department of Mathematics, Nagoya University,\\
Furo-cho Chikusa-ku, Nagoya 464-8602, Japan}
\affiliation[d]{Department of Mathematics, Shanghai University,\\
99 Shangda road, Shanghai 200444, China}

\emailAdd{shutaishigaki@shu.edu.cn, kinoshita@phys.chuo-u.ac.jp, matsumoto@shu.edu.cn}
\abstract{
	In this study, we analyze the dynamical stability of the D3-D7 model
	dual to a holographic conductor with a constant current under an external electric field.
	We particularly focus on the stability around the parameter region where the multivalued relation between the external electric field and the current is shown due to nonlinear conductivity.
	The dynamical stability of the states can be analyzed by considering linear perturbations in the background states and computing the quasinormal modes.
	In the multivalued region, we find that the states in one branch
	with a low electric current can be dynamically unstable.
	The turning point in the $J$--$E$ characteristic coincides with the stability switching.
	Further, we also find that the perturbations around the unstable states can become stable with finite wavenumber. 
	In other words, the perturbations in the background states become static at a critical wavenumber, implying the existence of inhomogeneous steady states with current filaments.
}

\begin{document}
\maketitle

\section{Introduction}
Systems far from equilibrium exhibit rich and interesting phenomena. 
Nevertheless, there are theoretically many difficulties in handling nonequilibrium systems, e.g. due to the non-conservation of energy, lack of detailed balance, and so on.
The gauge/gravity duality \cite{Maldacena:1998,Gubser:1998,Witten:1998} can be a powerful tool for studying nonequilibrium systems \cite{Liu:2018crr}.
The gauge/gravity duality allows us to study a specific, strongly-correlated field theory by mapping it to tractable dual gravity models in the appropriate limit even if the system is far from equilibrium.

Nonlinear electric conductivity is an interesting phenomena in
non-equilibrium physics.
In the context of the gauge/gravity duality, the D3-D7 model \cite{Karch:2002sh} shows nonlinear conductivity \cite{Karch:2007pd,Erdmenger:2007bn,Albash:2007bq} in the non-equilibrium steady state (NESS).%
\footnote{
	The Witten-Sakai-Sugimoto model also shows similar nonlinear conductivity \cite{Bergman:2008sg}.
}
This model consists of a stack of \(N_c\) D3-branes and 
\( N_f\) D7-branes. 
We will work in the large-\(N_c\) limit with the 't Hooft coupling fixed as \(\lambda = g_{\rm YM}^{2} N_{c}\), and the strong coupling $\lambda \gg1$. The D3-branes can be considered as the source of supergravity fields in AdS\(_5\times \mathrm{S}^5\) in the gravity picture.
In the dual field theory picture, open stings on the $N_c$
D3-branes correspond to the \(\mathcal{N}=4\) supersymmetric
$\mathrm{SU}(N_c)$ Yang-Mills (SYM) theory, and strings stretching
between the D7-branes and D3-branes correspond to \(\mathcal{N}=2\)
hypermultiplets.
In the probe limit \(N_c \gg N_f\), we can treat the D7-branes as probes and ignore the backreaction on the supergravity fields.
The dual field theory contains quark-like charged particles because the theory has a global \(\mathrm{U}(1)_{\text{B}}\) symmetry, which corresponds to the baryon charge. Thus, we can consider a system with a constant \(\mathrm{U}(1)_{\text{B}}\) current $J$ by applying an external electric field $E$.
In this study, we call this NESS regime as a ``holographic conductor''.
When the system has a finite baryon density, the charge is mainly carried by the background baryon density.
Notably, the conductivity can remain finite even if the baryon density vanishes.
In this case, the conduction is carried by the pair creation of the charged particle and antiparticle under the external electric field.
This is a holographic realization of the Schwinger effect and a many-body quantum effect in a non-equilibrium system \cite{Hashimoto:2013mua}.
Owing to the nonlinear conductivity in the D3-D7 model, the current can be multivalued with respect to the electric field \cite{Nakamura:2010zd}, meaning that there coexist two or more
solutions of the probe D$7$-brane for given parameters in the multivalued regime.

Now the question is which state should be physically realized in the dual field theory?
Similar multivalued behavior has been observed in the relation
between the quark condensate and temperature in the D3-D7 model at
\(E=0\) \cite{Mateos:2007vn}, which is a thermal equilibrium system.
Although one can discuss thermodynamic stability according to
thermodynamics in equilibrium systems, there is no well-established
guiding principle of thermodynamic stability in nonequilibrium systems
such as NESS. 
Meanwhile, it is a robust analysis to elucidate the dynamical stability
of perturbed systems even in nonequilibrium systems.
In the gauge/gravity duality, this is achieved by considering the corresponding field perturbations in the gravity side.%
\footnote{
	The dynamical stability of the D3-D7 model at finite temperature with finite density and \(E=0\) has been studied in \cite{Kaminski:2009ce} in the same way.
}
By imposing proper boundary conditions, we obtain quasinormal modes
(QNMs) with complex frequencies
\cite{Horowitz:1999jd,Nunez:2003eq,Kovtun:2005ev}. The complex
frequencies correspond to the poles of the retarded Green's function of
the corresponding excitations.
If there is a growing mode with a positive imaginary-part of the mode
frequency, the background state is dynamically unstable against the perturbations.

\begin{figure}[htbp]
\centering
\includegraphics[width=10cm]{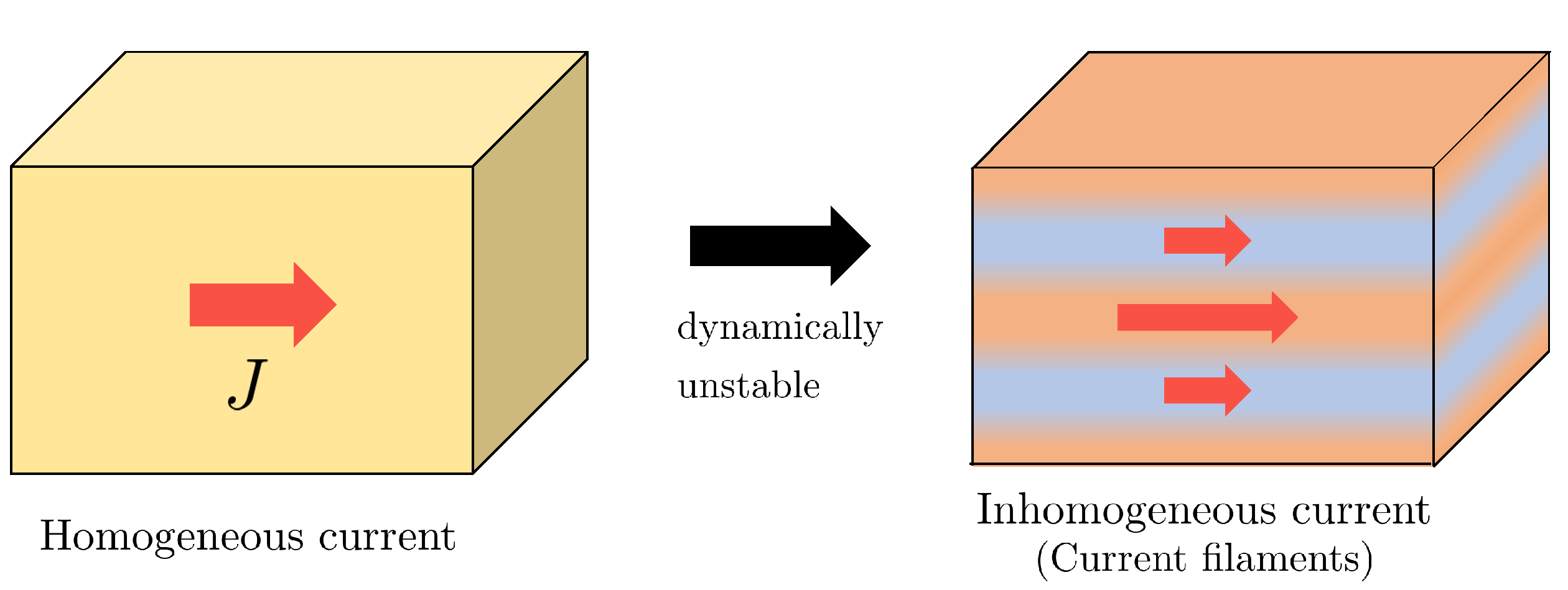}
\caption{The schematic of a new branch around the
 background states in the multivalued region. The system with the
 homogeneous current becomes dynamically unstable with respect to
 perturbations. For a finite wavenumber, we find a static perturbation
 on the background system. This indicates that the system with the
 spatially modulated current\,(current filaments) can be realized.}
\label{fig:intro}
\end{figure}

In this study, we investigate the dynamical stability of the holographic
conductor, particularly, in a multivalued regime.
We will consider perturbations of the current operator corresponding to
the worldvolume gauge field on D7-branes.
We focus on the perturbation of the gauge field parallel to the external
electric field $E$.
Notably, this perturbation should be coupled with a
perturbation of the brane embedding functions, implying an operator mixing between the corresponding operators: the current and quark condensate.
We numerically solved the coupled equations of motion by employing the determinant method according to ref.~\cite{Kaminski:2009dh}.
As a result, we found the dynamical instability
in the multivalued region, having multiple states with high and
low currents $J$ for a given electric field $E$ and
temperature $T$;
the low $J$ states tend to become unstable against homogeneous perturbations.
In addition, by analyzing inhomogeneous perturbations with finite
spatial wavenumbers around the unstable states, we find the
existence of a critical wavenumber such that the mode frequency becomes
zero, i.e., a marginally stable static perturbation.
This implies that nonlinear inhomogeneous states can emerge with the current density modulated.
Such an inhomogeneous current, the so-called {\it current filaments}, is seen in materials exhibiting S-shaped negative differential conductivity \cite{Schoell:2001}.
Figure \ref{fig:intro} shows the schematic of the dynamical instability of the homogeneous system and the realization of the spatially inhomogeneous states. 
In this sense, the dynamical instability we found can be considered the {\it filamentary instability} in the NESS system.

The reminder of this article is organized as follows.
In section~\ref{sec:Background}, we review the background solutions of the
probe D$7$-brane, which can realize the nonlinear conductivity in the
D$3$-D$7$ brane model. 
We reproduce the $J$--$E$ characteristics as multivalued functions at zero and finite temperatures.
We also show the quark condensate and effective temperature as functions of the electric field and temperature.
In section~\ref{sec:Perturbation}, we discuss the equations for the
linear perturbations on background fields and briefly review the
determinant method to find the QNMs.
In section~\ref{sec:Results}, we show the numerical results for the linear stability of the background states in the multivalued region.
In section~\ref{sec:conclusion_and_discussion}, we present conclusion and discussion.

\section{Background}\label{sec:Background}
In this section, we briefly review the holographic conductor in the
D3-D7 model with no baryon charge density. 
We show the phase diagram for the conducting and insulating
phases with respect to the external electric field and temperature of the heat bath. 
We also discuss the multivaluedness of the current and quark condensate for a given electric field and temperature.
In addition, we show the behavior of the effective temperature near the
phase boundary. 

\subsection{D3-D7 model}
In this study, we employ the D3-D7 model as a gravity dual of the many-body systems with charged particles at a finite temperature.
The $10$-dimensional background geometry is Schwarzschild-AdS\(_5 \times \mathrm{S}^5\):%
\footnote{
	For simplicity, we have set each radius of AdS$_5$ and $S^5$ parts to $1$.
}
\newcommand{\uh}{u_{\mathrm{H}}}
\begin{equation}
		\dd s^{2}_{10}= \frac{1}{u^{2}}\left(-f(u)\dd t^{2}+\dd\vec{x}^{2}+\frac{\dd u^{2}}{f(u)}\right)+\dd\Omega_{5}^{2},
	\label{eq:metric}
\end{equation}
where
\begin{equation}
	f(u)=1-\frac{u^{4}}{\uh^{4}}.
\end{equation}
Here, $(t,\vec{x})=(t,x,y,z)$ represent the coordinates of the dual gauge theory in the (3+1)-dimensional spacetime and $u$ denotes the radial coordinate of the AdS direction. 
In this coordinate system, the black hole horizon is located at $u=u_{\text H}$ and the AdS boundary is located at $u=0$. The Hawking temperature is given by $T=1/(\pi \uh)$, which corresponds to the heat bath temperature in the dual field theory. The metric of the $\mathrm{S}^{5}$ part is given by
\begin{equation}
	\dd\Omega_{5}^{2}=\dd\theta^{2}+\sin^{2}\theta \dd\psi^{2}+\cos^{2}\theta \dd\Omega_{3}^{2},
\end{equation}
where $\dd\Omega_{3}^{2}$ denotes the line element of the $\mathrm{S}^{3}$ part.
In this study, we introduce a single D7-brane ($N_{f}=1\ll N_{c}$), which fills the AdS$_{5}$ part and wraps the $\mathrm{S}^{3}$ part of the $\mathrm{S}^{5}$. The configuration of the D7-brane is determined by the embedding functions of $\theta$ and $\psi$.

The action for the probe D7-brane is given by the Dirac-Born-Infeld~(DBI) action, as follows:
\begin{equation}
	S_{\text{D7}}=-T_{\text{D7}} \int \dd^{8}\xi \sqrt{-\det \left(g_{ab}+(2\pi\alpha')F_{ab} \right)}
	+ S_{\text{WZ}},
\end{equation}
where \(S_{\text{WZ}}\) denotes the Wess-Zumino term but can be ignored in our setup. 
The D7-brane tension is given by 
$T_{\text D7}=(2\pi)^{-7}(\alpha')^{-4}g_{\text s}^{-1}$.
$g_{ab}$  is the induced metric defined as follows:
\begin{equation}
	g_{ab} = \frac{\partial X^{\mu}}{\partial \xi^{a}} \frac{\partial X^{\nu}}{\partial \xi^{b}}G_{\mu\nu},
\end{equation}
where $\xi^{a}$ denotes the worldvolume coordinate on the D7-brane with
$a,b=0,\cdots,7$ and $X^{\mu}$ denotes the target space coordinate with
$\mu,\nu=0,\cdots,9$. $G_{\mu\nu}$ represents the background metric given by
eq.~(\ref{eq:metric}). The field strength of the U(1) gauge field on the
D7-brane is given by $2\pi\alpha' F_{ab} \equiv \partial_{a}A_{b}-\partial_{b}A_{a}$.
In this study, we assume the following ansatz for fields:
\begin{equation}
	\theta=\theta(u), \hspace{1em} \psi =0, \hspace{1em} A_{x}=-Et + h(u).
\end{equation}
The induced metric and the field strength are respectively given by 
\begin{equation}
 g_{ab}\dd\xi^a\dd\xi^b = \frac{1}{u^{2}}
  \left(-f(u)\dd t^{2} + \dd\vec{x}^{2}\right)
  + \left(\frac{1}{u^2f(u)} + \theta'(u)^2\right)\dd u^{2}
  + \cos^2\theta(u) \dd\Omega_{3}^{2} ,
\end{equation}
and 
\begin{equation}
 2\pi\alpha' F_{ab}\dd\xi^a\wedge\dd\xi^b = - \frac{E}{2}\dd t\wedge \dd x
  - \frac{h'(u)}{2} \dd x \wedge \dd u ,
\end{equation}
where the prime denotes the derivative with respect to $u$.
Then, the DBI action can be expressed as follows:
\begin{equation}
 \begin{aligned}
  S_{\text D7} &= {\cal N}\int \dd t \dd^{3}x \dd u \, \mathcal{L} , \\
  \mathcal{L} &\equiv - \cos^{3}\theta(u) 
  g_{xx} \sqrt{- g_{tt}g_{xx}g_{uu}- g_{uu}E^{2} - g_{tt} h'(u)^{2}},
 \end{aligned}
 \label{eq:L_DBI}
\end{equation}
where the prefactor of the action is given by
\begin{equation}
	{\cal N}= T_{\text D7} (2\pi^{2}) =\frac{N_{c}\lambda}{(2\pi)^{4}} ,
\end{equation}
with the relation $4\pi g_{s} N_{c} \alpha'^{2}=1$ and 't Hooft coupling
$\lambda = g_{\text YM}^2 N_{c} = 4\pi g_s N_{c}$. 
Near the AdS boundary ($u=0$), the fields can be expanded as follows:
\begin{equation}
\begin{aligned}
	\sin\theta(u) &= m u + c u^3 + \cdots,\\
	A_x(t,u) & = - E t + h(u) = - E t + \frac{J}{2} u^2 + \cdots,
\end{aligned}
\label{eq:asymptotic}
\end{equation}
where, according to the AdS/CFT dictionary, the coefficients $m$ and $c$ are related to quark mass and quark
condensate~\cite{Mateos:2007vn}, and $E$ and $J$ are related to electric field and
electric current density~\cite{Karch:2007pd} in the dual field theory: 
\begin{equation}
 m_q = \frac{\lambda^{1/2}}{2\pi} m ,\quad
  \left<\bar{q}q \right> = 2 \frac{N_c}{(2\pi)^3} \lambda^{1/2} c ,
\end{equation}
and 
\begin{equation}
 \mathcal{E}_x = \frac{\lambda^{1/2}}{2\pi} E ,\quad
  \left<\bar{q} \gamma^x q \right> = \frac{N_c}{(2\pi)^3}\lambda^{1/2} J .
\end{equation}
Hereinafter, for simplicity, we will refer to dimensionless, geometrical parameters
$m$, $c$, $E$, and $J$ as quark mass, quark condensate, electric field, and
electric current, respectively.
Because the D$7$-brane action does not explicitly contain $h(u)$ but
depends only on $h'(u)$, the following quantity 
\begin{equation}
 \frac{\partial\mathcal{L}}{\partial h'} = 
  \frac{\cos^3\theta g_{tt}g_{xx} h'}
  {\sqrt{(- g_{tt}g_{xx} - E^{2})g_{uu} - g_{tt} h'^{2}}} 
  \label{eq:conserved_quantity}
\end{equation}
is conserved with respect to $u$-derivative.
Substituting the asymptotic behavior (\ref{eq:asymptotic})
near the AdS boundary, we find that this conserved quantity coincides
with the electric current density as 
$J = -\lim_{u\to 0} \partial\mathcal{L}/\partial h'$.
Meanwhile, in the denominator of (\ref{eq:conserved_quantity}) the term 
$-g_{tt}g_{xx} - E^2$ can go to zero because $-g_{tt}$ will decrease
toward the horizon of the bulk black hole at $u=u_{\text H}$.%
\footnote{
In zero temperature cases ($u_{\text H} \to \infty$), 
this horizon will be identical with the Poincar\'e horizon of pure AdS.
}
At the locus $u=u_* \equiv u_{\text H}/(1 + u_{\text H}^4 E^2)^{1/4}$ 
such that $-g_{tt}g_{xx} - E^2 = 0$, 
we have 
\begin{equation}
 \left.\frac{\partial\mathcal{L}}{\partial h'}\right|_{u=u_*}
  = - \cos^3\theta(u_*) \sqrt{-g_{tt}}g_{xx} = -J.
  \label{eq:current_density}
\end{equation}
Thus, the electric current density $J$ is determined locally at $u=u_*$.
In fact, the locus $u=u_*$ is referred to as the `` effective horizon'' 
on the worldvolume of the D$7$-brane because a causal boundary is given by an effective metric on the
worldvolume, which governs dynamics of brane.%
\footnote{This fact plays a significant role in studying the time evolution
of brane dynamics. For example, see ref.~\cite{Hashimoto:2014yza}.}
For the DBI action, the effective metric is the so-called open-string
metric defined by 
$\gamma_{ab} \equiv g_{ab} + (2\pi\alpha')^2 F_{ac}F_{bd}g^{cd}$
~\cite{Kim:2011qh,Seiberg:1999vs}:
\begin{equation}
 \begin{split}
  \gamma_{ab}\dd\xi^a\dd\xi^b 
  &= \frac{g_{tt}g_{xx} + E^2}{g_{xx}} \dd t^2
  - 2\frac{Eh'}{g_{xx}} \dd t \dd u + 
  \frac{g_{xx}g_{uu} + h'^2}{g_{xx}} \dd u^2 \\
  &\quad + \left(g_{xx} + \frac{E^2}{g_{tt}} + \frac{h'^2}{g_{uu}}\right) \dd x^2
  + \frac{\dd y^2 + \dd z^2}{u^2} + \cos^2\theta(u) \dd\Omega_3^2 .
 \end{split}
 \label{eq:effmetric}
\end{equation}
This shows that the locus $u=u_*$ is the Killing horizon with respect
to time coordinate $t$ on the AdS boundary, which corresponds to the time in
the dual field theory.

\subsection{D7-brane embeddings}
To obtain the configuration of the D$7$-brane, we should solve
the equations of motion for $\theta(u)$ and $h(u)$, which are given by
nonlinear second-order ordinary differential equations.
There are three types of solutions depending on
the bulk behavior: the so-called {\em Minkowski}, 
{\em black hole}, and {\em critical embeddings}~\cite{Mateos:2006nu,Mateos:2007vn,Frolov:2006tc}.
The solutions can be classified according to whether a horizon emerges on the worldvolume of the
D$7$-brane.

First, for the Minkowski embeddings, the solutions of the brane embedding
function $\theta(u)$ can smoothly reach $\theta(u_0)=\pi/2$ ($u_0<u_*$), 
at which the size of
$S^3$ wrapped by the D$7$-brane shrinks to zero, without encountering any horizon on
the brane.
In this phase, fluctuations on the D$7$-brane are confined and cannot be dissipated, meaning that they are generally described by normal modes with real
frequencies, which correspond to stable excitations of meson, i.e.,
quark/antiquark bound states in the dual field theory~\cite{Kruczenski:2003be}.%
\footnote{
	A degree of freedom of messino corresponds to the fermionic fields in the D3-D7 model what we ignored in this paper.
	In a massless case, it has been studied in ref.~\cite{Kirsch:2006he}.
	Recently,  a spectrum of the messino in a massive case has been studied in 
	ref.~\cite{Abt:2019tas}.
}
In addition, we can see that the electric current density vanishes even
under external electric fields.
Thus, we can consider the Minkowski embedding phase as the insulating
state in the dual field theory.

Second, for the black hole embeddings, the effective horizon emerges at 
$\theta(u_*)<\pi/2$.
In this phase, fluctuations on the brane can eventually dissipate into
the effective horizon, described by QNMs with
complex frequencies, implying that the excitations of meson become unstable, and then
quark/antiquark bound states will dissociate in the dual field theory.
Under external electric fields, the current density $J$ becomes finite
as previously mentioned in eq.~(\ref{eq:current_density}).
Hence, we can consider the black hole embedding phase as the conducting
state in the dual field theory. 
In particular, because the current density with Joule heating does not depend on time, 
the system is the NESS with a
constant electric current.

Finally, the critical embeddings are critical solutions between the
Minkowski and black hole embeddings. 
The embedding function $\theta(u)$ reaches $\theta=\pi/2$ at $u=u_*$,
and the D$7$-brane has a conical configuration with no current density, $J=0$.
(For example, see ref.~\cite{Hashimoto:2015wpa}.)

Depending on which type of solution we wish to obtain, we should impose the appropriate boundary condition at $u=u_{0}$ or $u=u_{*}$. 
Near the AdS boundary $u=0$, we can read the observables in the dual field
theory according to eq.~(\ref{eq:asymptotic}).

\subsection{Phase diagram}
\begin{figure}
\centering
\includegraphics[width=10cm]{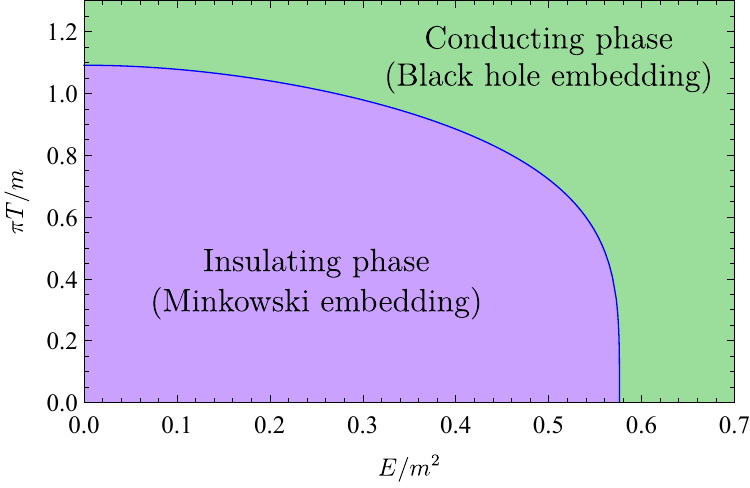}
\caption{The phase diagram of the D3-D7 model in 
the electric field and temperature plane.
The solid blue curve denotes the boundary between the black hole embedding\,(green region) and Minkowski embedding\,(purple region), which is called the critical embedding. These two embeddings correspond to the conducting phase and insulating phase in the dual field theory, respectively.}
\label{fig:phase}
\end{figure}

In our case, a family of solutions depends on three 
parameters, $T$, $E$, and $m$.
The system is invariant under the following scale transformation, $u \to k u$, 
$(t,\vec{x}) \to (kt,k\vec{x})$, $u_{\text H} \to k u_{\text{H}}$, 
$m \to m/k$, $c \to c/k^3$, $E \to E/k^2$, $J \to J/k^3$, for arbitrary
constant $k$.
Without loss of generality, we can characterize the system by two scale-free parameters
$ \left( \pi T/m, E/m^{2}  \right) $, 
which are scaled by the quark mass.
Figure \ref{fig:phase} shows the phase diagram of the D3-D7 model.
The solid blue curve denotes the critical embeddings between the black
hole embeddings and Minkowski embeddings.
The origin $E=T=0$ is the vacuum state in the dual field theory~\cite{Karch:2002sh}, which is described by an exact 
solution of the embedding function $\theta(u) = \arcsin mu$.
From figure \ref{fig:phase}, the conducting state, which corresponds to the black hole embedding, is realized at a high temperature or large electric field.
As we move to the top, the dissociation of the mesons
is caused by the thermal effect of the gluon heat bath.
At $E=0$, for example, the critical temperature is given by $\pi T_{\rm crit}/m= 1.09068$.
Meanwhile, as we move to the right, the dissociation of the mesons is
caused by the external electric field. 
The former process corresponds to a ``meson melting'' discussed
in ref.~\cite{Mateos:2007vn}, whereas the latter corresponds to dielectric breakdown \cite{Erdmenger:2007bn, Albash:2007bq}.
Moreover, the high temperature or large electric field limit
corresponds to a limit in which the quark is massless $m=0$.
In this case, the brane embedding can be trivial, 
$\theta(u)=0$, and the current is analytically given by 
$J=E(\pi^4 T^4 + E^2)^{1/4}$.

\begin{figure}
\centering
\includegraphics[width=7cm]{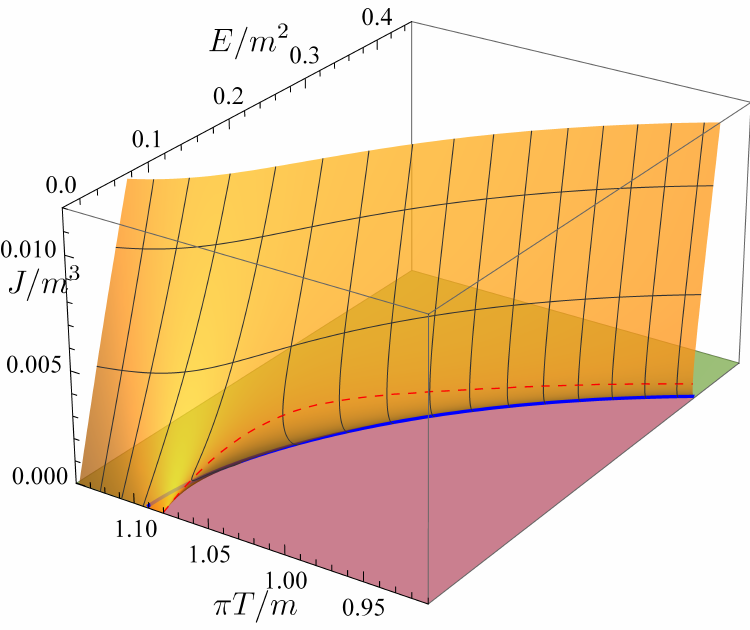}
\includegraphics[width=7cm]{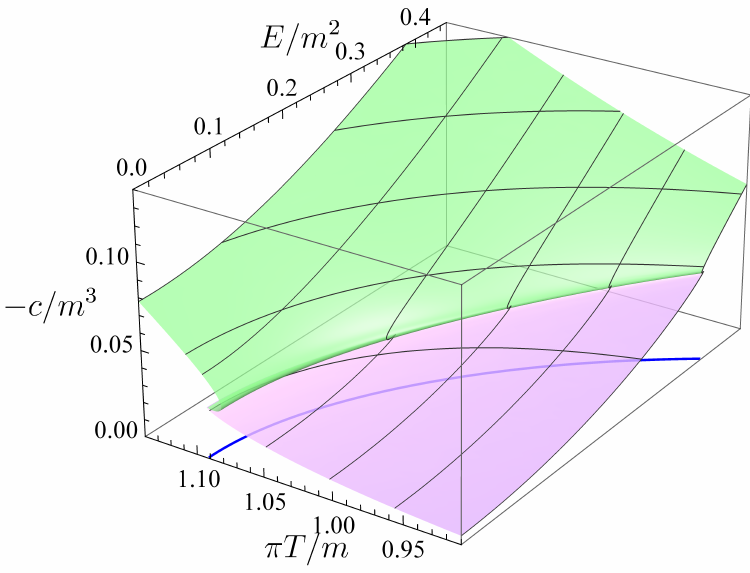}
\caption{The left panel shows the plot of $J/m^{3}$ with respect to $E/m^{2}$ and
 $\pi T/m$ near critical temperature 
in the phase diagram. The purple and green regions denote the
 Minkowski and black hole embeddings, respectively. The blue curve
 denotes the critical embedding as shown in figure \ref{fig:phase}. The
 vertical black curve shows the $E$--$J$ curves for each fixed $\pi T/m$. The red dashed curve corresponds to the ``turning point'' of each $E$--$J$ curve. The right panel shows the plot of $-c/m^{3}$ with respect to $E/m^{2}$ and $\pi T/m$ in the same parameter region of the left panel. The purple and green regions again correspond to the Minkowski and black hole embeddings, respectively. We focus on the range of $0< E/m^{2} \lesssim 0.45$ and $0.91 \lesssim \pi T/m \lesssim 1.14$.}
\label{fig:ETJ}
\end{figure}

The phase diagram in the vicinity of the critical
embeddings shows a self-similar
structure~\cite{Mateos:2006nu,Mateos:2007vn,Frolov:2006tc}.
Because of this critical behavior, the black hole and Minkowski
embeddings will coexist for the same control parameters around a phase
boundary represented by the critical embeddings.
In fact, it turns out that the current density is multivalued as a
function of the electric field in the $E$--$J$ curve \cite{Nakamura:2010zd}.
Thus, we show the plot of 
electric current $J/m^{3}$
and quark condensate $-c/m^{3}$
with respect to $E/m^{2}$ and $\pi T/m$ in figure \ref{fig:ETJ}. 
We focus on the phase diagram around the phase
boundary, $0.91 \lesssim \pi T/m \lesssim 1.14$.
In the left panel of figure \ref{fig:ETJ}, the purple and green regions at the bottom denote the Minkowski and black hole embeddings, respectively. 
The vertical black curves on the surface denote $E$--$J$ curves for fixed temperatures $\pi T/m$.
We can find that some curves obviously turn over as $J$ decreases, i.e., the sign of $(\partial J/\partial E)_{T,m}$ changes.
The red dashed curve on the surface corresponds to the ``turning point'' of each $E$--$J$ curve.
Notably, at $E=0$, the red dashed curve ends at the point ($\pi T/m = 1.08024$) below the
critical embedding\,(shown by the endpoint of the blue curve:~$\pi T_{\rm crit}/m = 1.09068$). This agrees with the phase structure in the D3-D7 model with
$E=0$\,\cite{Mateos:2007vn}. That is, the black hole embedding can be
found below the value of $\pi T/m$ at the critical embedding.
This can be confirmed from the right panel of figure \ref{fig:ETJ}. One can find the black hole embedding solutions, which correspond to each point on the green surface in the right panel of figure \ref{fig:ETJ}, below $\pi T_{\rm crit}/m$.
If the brane configuration is the black hole embedding in
the absence of the electric field, applying the electric field can
instantly produce the current.
Therefore, in the high-temperature region ($\pi T/m \geq 1.08024$), we can observe linear conduction
obeying Ohm's law.

\begin{figure}
\centering
\includegraphics[width=7cm]{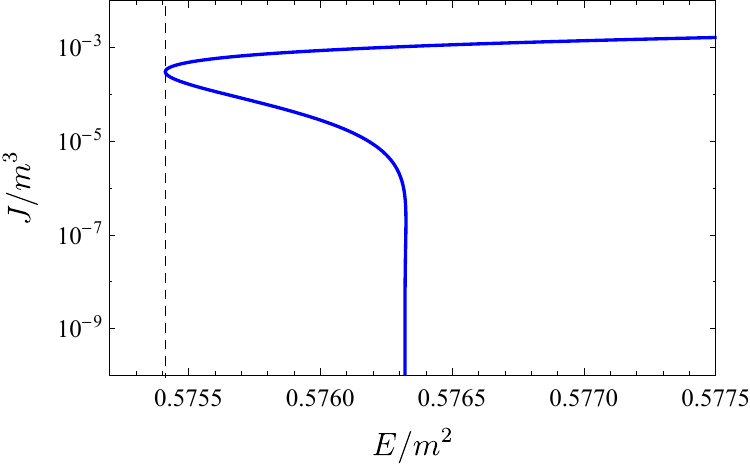}
\includegraphics[width=7cm]{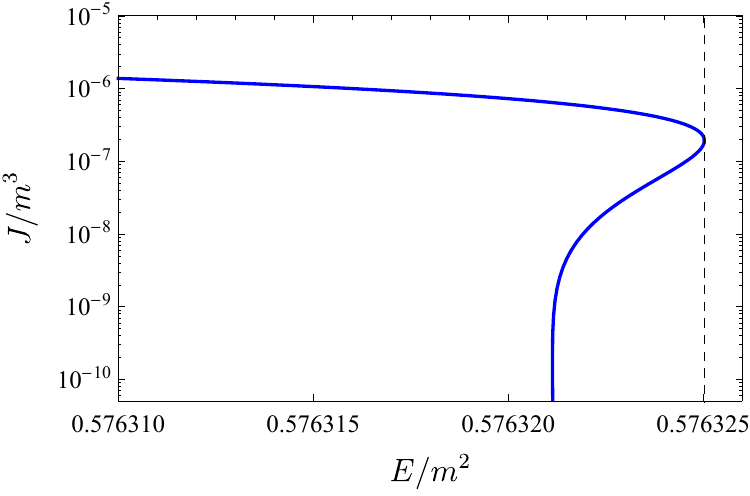}
\caption{
	The $E$--$J$ curve for $T=0$. The vertical axis is shown in
	a log scale. The right panel shows the zooming in of the left panel
	around the critical value of the electric field. The vertical dashed
	lines in the left panel and the right panel denote the electric field at the first and second turning point values, respectively:
	$(E^{(1)}_{\rm turn}/m^{2},\, J^{(1)}_{\rm turn}/m^{3})= (0.575412,\, 3.0031\times10^{-4})$ and
	$(E^{(2)}_{\rm turn}/m^{2},\, J^{(2)}_{\rm turn}/m^{3})= (0.5763250,\, 1.9178\times10^{-7})$.
}
\label{fig:zeroT}
\end{figure}

From each $E$--$J$ curve (the vertical black curves), the $J/m^{3}$ is multivalued with respect to $E/m^{2}$.
At first glance, $J/m^{3}$ seems to be the two-valued function of $E/m^{2}$ in the left panel of figure \ref{fig:ETJ}. 
However, the self-similar behavior around the critical embedding yields
more complicated structure.
In the vicinity of the critical embedding, we observe that the current density decreases again with electric field. In figure \ref{fig:zeroT}, for instance, we show the behavior near the critical embedding at $T=0$.
For later convenience, we specify the first and second turning point values
$(E^{(1)}_{\rm turn}/m^{2},\, J^{(1)}_{\rm turn}/m^{3})= (0.575412,\, 3.0031\times10^{-4})$
and
$(E^{(2)}_{\rm turn}/m^{2},\, J^{(2)}_{\rm turn}/m^{3})= (0.5763250,\, 1.9178\times10^{-7})$
, respectively, as denoted by the vertical dashed line in figure \ref{fig:zeroT}.
As shown in figure \ref{fig:zeroT}, the second turning point in the
$E$--$J$ curve appears for a small value of $J/m^{3}$. 
Although we have checked up to the second turning point within numerical
accuracy, we expect that this turning structure will appear repeatedly
in the vicinity of the critical embedding. 

\subsection{Effective temperature}
\begin{figure}
\centering
\includegraphics[width=0.6\linewidth]{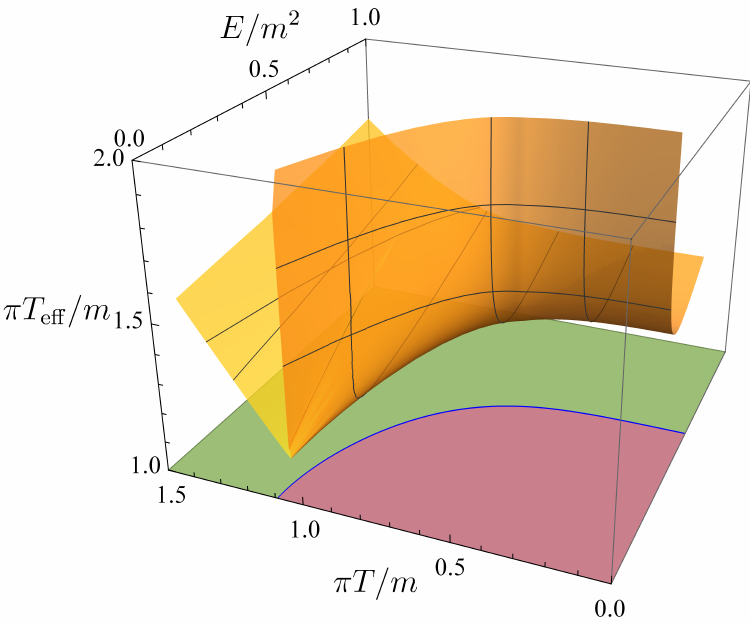}
\caption{The plot of \(\pi T_{\text{eff}}/m\) with respect to \(E/m^2\) and \(\pi T/m\) within \(0 < E/m^2< 1\) and \(0< \pi T/m < 1.5\). The purple and green regions and solid blue curve on the bottom denote the Minkowski embedding, black hole, and critical embeddings, respectively.}
\label{fig:effT}
\end{figure}

In the presence of the electric field, the dynamics of the D7-brane are governed by the effective metric (\ref{eq:effmetric}). The surface gravity on the effective horizon $u=u_{*}$ is defined by 
\begin{equation}
	\kappa= - \left. \sqrt{\frac{-\gamma_{tt}}{\gamma_{tu}^{2}-\gamma_{tt}\gamma_{uu}}}\frac{\dd}{\dd u}\sqrt{-\gamma_{tt}} \right|_{u=u_{*}}
	= -\frac{\gamma_{tt}'(u_{*})}{2 \gamma_{tu}(u_{*})}
	=\frac{2}{u_{*}^{5}E h'(u_{*})}.
	\label{eq:surface_gravity}
\end{equation}
In the middle expression, the minus sign comes from \(\gamma_{tu}<0\).
Then, one can define the effective Hawking temperature as $T_{\rm eff}\equiv \kappa/(2\pi)$. Figure \ref{fig:effT} depicts the plot of $\pi T_{\rm eff}/m$ with respect to $E/m^{2}$ and $\pi T/m$.
To interpret the behavior of the effective temperature, we solve eq.~(\ref{eq:conserved_quantity}) with respect to $h'(u)$ and obtain
\begin{equation}
	h'(u)^{2}= \frac{J^{2}(-E^{2}-g_{tt}g_{xx})g_{uu}}{g_{tt}\left(J^{2} +\cos^{6}\theta g_{tt}g_{xx}^{2} \right)}.
\end{equation}
Because both the numerator and denominator of $h'(u)$ become zero simultaneously in the limit of $u\to u_{*}$, $h'(u_{*})$ is given by applying L'Hospital's rule,
\begin{equation}
	h'(u_{*})^{2} = \left. -\frac{(-g_{tt}g_{xx})'}{(\cos^{6}\theta g_{tt}g_{xx}^{2})'}\cos^{6}\theta g_{xx}^{2}g_{uu} \right|_{u=u_{*}}.
\end{equation}

At low temperature, 
in the critical embedding, the current density, $J$, becomes zero, whereas
the electric field, $E$, remains a finite value, which corresponds to $\theta(u_{*})\to \pi/2$ from eq.~(\ref{eq:current_density}). Thus, $h'(u_{*}) = 0$ with finite $E$, and the effective temperature diverges at the critical embedding (figure \ref{fig:effT}).
Meanwhile, at high temperature, if we can take the limit of $E\to 0$ (identical to $u_{*}\to u_{\rm H}$) in the black hole embedding, we obtain
\begin{eqnarray}
	 E^{2}h'(u_{*})^{2} &=& \left. \frac{(-g_{tt}g_{xx})'}{(\cos^{6}\theta g_{tt}g_{xx}^{2})'}\cos^{6}\theta g_{tt}g_{xx}^{3}g_{uu} \right|_{u=u_{*}} \nonumber\\
	 &\to& -\left. g_{tt}g_{xx}^{2}g_{uu} \right|_{u=u_{\rm H}}
	  \hspace{1em} (u_{*}\to u_{\rm H}) ,
\end{eqnarray}
where we used $g_{tt}\to 0$ in the limit of $u_{*}\to u_{\rm H}$. Therefore, we confirm that $\pi T_{\rm eff} \to 1/u_{\rm H} =\pi T$, i.e., the effective temperature becomes identical to the heat bath temperature at $E=0$. In figure \ref{fig:effT}, we can see this behavior in the linear conduction regime at high temperatures.
As a result, the effective temperature clearly shows the qualitative difference between dielectric breakdown and linear conduction.

\section{Linear perturbations}\label{sec:Perturbation}

In this section, we formulate linear perturbations on the probe D7-brane.
The general ansatz for perturbations of the embedding functions and the
worldvolume gauge fields on the D$7$-brane is given by 
\[
	X^{\mu} \to X^{\mu} + \epsilon \delta X^{\mu},~~
	A_a \to A_a + \epsilon \delta A_a,
\]
where $X^\mu$ and $A_a$ are background solutions, and $\epsilon$ is a small bookkeeping parameter.
Expanding the DBI action to quadratic order in $\epsilon$, we obtain the effective action%
\footnote{
	See appendix A in ref.~\cite{Mas:2008jz}.
	Eq.~(A.9) is the corresponding Lagrangian density.
}
for the perturbations as follows:
\begin{equation}
	S^{(2)} = -\frac{T_{\text{D7}}}{2}\int \dd^8\xi
	\sqrt{-M}
\left[
		M^{ab}\mathcal{M}^{(2)}_{ba}
		+ \frac{1}{4} (M^{ab}\mathcal{M}^{(1)}_{ba})^2
		- \frac{1}{2} M^{ab}\mathcal{M}^{(1)}_{bc} M^{cd}\mathcal{M}^{(1)}_{da}
	\right],
\end{equation}
where
\begin{subequations}
\begin{align}
	\mathcal{M}^{(1)}_{ab} =& 2 \partial_{[a} \delta A_{b]}
	+ \partial_a X^{\mu} \partial_b X^{\nu} \pdv{G_{\mu\nu}}{X^{\rho}} \delta X^{\rho}
	+ 2 G_{\mu\nu} \partial_{(a} X^{\mu} \partial_{b)} \delta X^{\nu},\\
	\mathcal{M}^{(2)}_{ab} = &
	G_{\mu\nu} \partial_a \delta X^{\mu}\partial_b \delta X^{\nu}
	+ 2 \pdv{G_{\mu\nu}}{X^{\rho}} \partial_{(a} X^{\mu} \partial_{b)} \delta X^{\nu} \delta X^{\rho} \nonumber\\
	&+  \frac{1}{2}\partial_a X^{\mu} \partial_b X^{\nu} \pdv{G_{\mu\nu}}{X^{\rho}}{X^{\kappa}}\delta X^{\rho} \delta X^{\kappa},
\end{align}
\end{subequations}
and \(M_{ab} \equiv g_{ab} + F_{ab}\).
Notably, $M^{ab}$ is defined by the inverse of $M_{ab}$ such that
$M^{ac}M_{cb} = M_{bc}M^{ca} = \delta^a{}_b$ is satisfied.


For our purpose of studying the linear stability of the multivalued background profile, we consider the following ansatz for perturbed fields:
\begin{equation}
\begin{aligned}
	A_{a} \dd \xi^a \to& A_x \dd x + \epsilon \delta A_{x}(t,\vec{x}_{\perp},u)\dd x,\\
	\theta \to& \theta(u) + \epsilon \delta \theta(t,\vec{x}_{\perp},u),
\end{aligned}
\end{equation}
where $\vec{x}_{\perp} = (y,z)$ is perpendicular to the $x$-direction.
We have assumed that the perturbations do not depend on the $x$-coordinate
parallel to the background electric field and current.%
\footnote{
	When $\delta A_x$ depends on $x$, we have to contain the perturbation field $\delta A_t$, additionally.
}
Because any other perturbations are decoupled from the above
perturbations, we can set the other perturbations to zero without loss
of generality.
Thus, the quadratic action can be expressed as follows:
\begin{equation}
	S^{(2)} = -\frac{\mathcal{N}}{2} \int \dd{t} \dd[3]{\vec{x}} \dd{u} \left[
		(\partial_{\alpha} \tilde{\Phi}^{T}) \tilde{A}^{\alpha\beta} \partial_{\beta} \tilde{\Phi}
		+
		\tilde{\Phi}^{T} \tilde{B}^{\alpha} \partial_{\alpha} \tilde{\Phi}
		+
		\tilde{\Phi}^{T} \tilde{C} \tilde{\Phi}
	\right],
\end{equation}
where \(\tilde{A}^{\alpha \beta}, \tilde{B}^{\alpha}\) and \(\tilde{C}\) denote coefficient matrices.
Each component of the matrices is given by eqs.~(\ref{eq:coeffs_position}).
The indices \(\alpha, \beta\) denote coordinates \(t, x, y, z, u\).
We have written the perturbations as \(\tilde{\Phi} = [\delta \theta, \delta A_x]^T\) in the above equation.
Considering the Fourier transformation with respect to \(t\) and \(\vec{x}_{\perp}\), we can rewrite \(\tilde{\Phi}\) as follows:
\begin{equation}
	\tilde{\Phi}(t, \vec{x}_{\perp}, u)
	=
	\int \frac{\dd{\omega}\dd[2]{\vec{k}_{\perp}}}{(2 \pi)^3}
	e^{-i(\omega t - \vec{k}_{\perp} \cdot \vec{x}_{\perp})}
	\Phi(\omega, k_{\perp}, u) ,
\end{equation}
where $k_{\perp} \equiv |\vec{k}_{\perp}|$.
The Fourier components of the perturbations depend on $k_{\perp}$
because the background system is isotropic in the $(y,z)$-space.
Because our system and the background solutions are static and homogeneous, we can rewrite the quadratic action in the momentum space.
Writing \(\Phi_{\pm k} = \Phi(\pm\omega, k_{\perp},u)\), we obtain the quadratic action as follows:
\begin{equation}
\begin{aligned}
	S^{(2)} = - \frac{\mathcal{N}}{2}\int\frac{\dd{\omega}\dd[2]{\vec{k}_{\perp}}}{(2 \pi)^3}
	\int \dd{u}
	\Big[
		(\partial_u \Phi_{-k}^{T}) A_k \partial_u \Phi_{k}
		&+ \Phi_{-k}^{T} B_k \partial_u \Phi_{k}\\
		&+ (\partial_u \Phi_{-k}^{T}) B_k^{\dagger} \Phi_{k}
		+ \Phi_{-k}^{T} C_k \Phi_{k}
	\Big],
\end{aligned}
\label{eq:quadratic_action_momentum}
\end{equation}
where
\begin{subequations}
\begin{gather}
	A_k =R(u)\gamma^{uu}
	\begin{bmatrix}
		\Xi(u) & -M^{xu}\theta'(u) \\
		-M^{xu}\theta'(u) & \gamma^{xx}
	\end{bmatrix},\\
	B_k =R(u)
	\begin{bmatrix}
		i \omega \gamma^{tu} \Xi(u) - 3 \gamma^{uu}\theta'(u) \tan\theta&
		 i \omega M^{tx} \gamma^{uu}\theta'(u) - 3 M^{xu} \tan\theta \\
		-i\omega(M^{tx} \gamma^{uu} + 2\gamma^{tu}M^{xu}) \theta'(u) &
		 i\omega \gamma^{tu} \gamma^{xx}
	\end{bmatrix},\\
	C_k =R(u)
	\begin{bmatrix}
		\Xi(u)\Omega(u)^2  - 3 + 6 \tan^2\theta&
		-\Omega(u)^2 M^{xu} \theta'(u) - 3 i\omega M^{tx}\tan\theta\\
		-\Omega(u)^2 M^{xu} \theta'(u) + 3 i\omega M^{tx}\tan\theta&
		\Omega(u)^2 \gamma^{xx}
	\end{bmatrix}.
\end{gather}
\end{subequations}
We have defined
\begin{equation}
	\Xi(u) = 1- \gamma^{uu}\theta'(u)^2, ~~\Omega(u)^2 = \omega^2 \gamma^{tt} + k^2_{\perp} \gamma^{\perp\perp},
	\label{eq:Xi_Omega_definition}
\end{equation}
and $R(u) = - \mathcal{L}$ given by eq.~(\ref{eq:L_DBI}).
Notably, $\gamma^{ab}$ denotes the inverse of the effective metric
(\ref{eq:effmetric}), which is identical to $M^{(ab)}$.
In the present case we have $M^{xu} = - M^{ux}$ and $M^{tx} = -M^{xt}$.
Hereinafter, we simply write \(\Phi\) as \(\Phi_k\).
\(\Phi^{I}\) denotes each component of \(\Phi\) with \(I = 1,2\).
For convenience, we introduce a new variable \(\bar{\Phi}\) by
\begin{equation}
	\Phi^{I}(u) = \varrho(u) \bar{\Phi}^I(u), \quad
	 \varrho(u) = \mathrm{diag}(u, 1) ,
\end{equation}
to remove the factors of non-normalizable modes, $u^{\Delta_{-}}$,
from the perturbations in the vicinity of the AdS boundary.
We also write the coefficient matrices for \(\bar{\Phi}\) as follows:
\begin{subequations}
\begin{align}
	\bar{A}_k =& \varrho(u) A_k \varrho(u),\\
	\bar{B}_k = & \varrho(u) B_k \varrho(u) + \varrho'(u) A_k \varrho(u),\\
	\bar{C}_k = & \varrho(u) C_k \varrho(u)
	+ \varrho(u) B_k \varrho'(u) + \varrho'(u) B_{k}^{\dagger} \varrho(u)
	+ \varrho'(u) A_k \varrho'(u).
\end{align}
\end{subequations}
Then, the quadratic action for \(\bar{\Phi}\) is given by the same form as eq.~(\ref{eq:quadratic_action_momentum}).
The equations of motion derived from the action are written as follows:
\begin{equation}
	\partial_u
	\left(
		\bar{A}_{k} \partial_u \Phi
		+ \bar{B}_k^{\dagger} \Phi
	\right)
	- \bar{B}_{k} \partial_u \Phi
	- \bar{C}_{k} \Phi = 0.
	\label{eq:linear_eom_kspace}
\end{equation}
We solve eq.~(\ref{eq:linear_eom_kspace}) with proper boundary conditions in the context of the AdS/CFT correspondence.
As mentioned in the previous section, the effective horizon is located
at $u=u_*$
on the brane's worldvolume in the conducting phase.
Because the effective horizon is a characteristic surface for the
perturbation equations, an ingoing-wave boundary
condition should be imposed to study the responses to perturbations.%
\footnote{
	This choice of the boundary condition in the NESS setup has been employed and discussed in refs.~\cite{Mas:2009wf,Ishigaki:2020coe,Ishigaki:2020vtr}
}
At the effective horizon, we have $\gamma^{uu} = 0$, so that $u=u_*$ is a singular point for the perturbation equations. 
We consider the Frobenius expansion at \(u=u_*\), as follows:
\begin{equation}
	\bar{\Phi} (u) =
	\left(1 - \frac{u}{u_*}\right)^{i \lambda} \bar{\Phi}_{\text{reg}}(u),
	\label{eq:Frobeniu_expansion}
\end{equation}
where \(\bar{\Phi}_{\text{reg}}(u)\) denotes a regular function at \(u=u_*\)
and 
\(\lambda\) is a constant.
Substituting eq.~(\ref{eq:Frobeniu_expansion}) into eq.~(\ref{eq:linear_eom_kspace}), we obtain the characteristic equation for \(\lambda\):
\begin{equation}
	\lambda^2 (a \lambda^2 + b\lambda + c) = 0,
\end{equation}
where
\begin{equation}
\begin{gathered}
	a \equiv (\gamma^{uu})'(u_*)^2 d,~~
	b \equiv - 4 \omega \gamma^{tu} (\gamma^{uu})'(u_*) d,~~
	c \equiv -9 (M^{xu})^2 \tan^2\theta + 4 \omega^2(\gamma^{tu})^2 d,\\
	d \equiv \gamma^{xx} - (M^{xu})^2 \theta'(u_*)^2.
\end{gathered}
\label{eq:coeffi_characteristic}
\end{equation}
In the above equations, all functions are evaluated at \(u=u_*\).
Solving the equation for \(\lambda\), we find four roots:
\begin{equation}
	\lambda_{0} = 0,~~
	\lambda_{1} =
	\frac{2 \omega \gamma^{tu} \pm 3 M^{xu}\tan \theta /
	\sqrt{d}}{(\gamma^{uu})'(u_*)}
= \frac{\omega}{\kappa} \pm \frac{3 M^{xu}\tan \theta /
	\sqrt{d}}{(\gamma^{uu})'(u_*)} 
,
	\label{eq:Frobenius_exponents}
\end{equation}
where the former one is a double root of the characteristic equation.
In the above equation for \(\lambda_1\), we have used the relation between \(\gamma^{ab}\) and \(\kappa\) defined by eq.~(\ref{eq:surface_gravity}), as follows:
\begin{equation}
	\kappa = \frac{(\gamma^{uu})'(u_*)}{2\gamma^{tu}(u_*)}.
	\label{eq:surface_gravity_in_gamma-Inv}
\end{equation}
A choice of \(\lambda = \lambda_{0}\) corresponds to the ingoing-wave solutions in our setup.%
\footnote{
	This implies that the coordinates \((t,u)\) take a form of 
	ingoing Eddington-Finkelstein like coordinates at the effective horizon.
}
To assess this simply, we should consider the high-frequency limit.
For sufficiently large $\omega$, we can drop the second term for \(\lambda_1\) in eq.~(\ref{eq:Frobenius_exponents}),
and then \(\lambda_0\) and \(\lambda_1\) reduce to the Frobenius
exponents for the perturbation of the gauge field transverse to \(E\) in the same background 
obtained in ref.~\cite{Ishigaki:2020coe}.
In that case, the transverse perturbation is decoupled from any other perturbations.
It is easy to show that $\lambda=0$ and $\lambda=\omega/\kappa$
correspond to the ingoing- and outgoing-wave solutions in the tortoise coordinate of the effective metric, respectively. (See appendix B of ref.~\cite{Ishigaki:2020coe}.)
Thus, $\lambda_0=0$ is expected to be the ingoing-wave solutions in eq.~(\ref{eq:Frobenius_exponents}), too.
More rigorous derivation of the ingoing-wave conditions for our system
is presented in appendix~\ref{appendix:ingoing}.

We are interested in the stability of this system when we allow
fluctuations in the current for a fixed electric field.
For this purpose, we should explore the modes of the perturbations
$\bar{\Phi}(u)$ 
satisfying boundary conditions such that non-normalizable modes and
outgoing-wave solutions vanish
at the AdS boundary ($u=0$) and effective horizon ($u=u_*$), respectively.
Therefore, by setting \(\lambda=0\), we will determine solutions with a mode frequency
$\omega$ by imposing the conditions 
\begin{equation}
	\bar{\Phi}_{\text{reg}}(u=u_*) = \bar{\varphi},~~\bar{\Phi}_{\text{reg}}(u=0) = 0,
\end{equation}
where \(\bar{\varphi}\) represents a pair of some constant values.
The \emph{determinant method}~\cite{Amado:2009ts,Kaminski:2009dh} helps us find eigenfrequencies and eigensolutions for such a problem.
We consider for the given $\omega$ and $k_{\perp}$ a set of linearly-independent, ingoing-wave
solutions $\{\phi_{(J)}(u)\}$ labeled by \(J\).
In our computation, we set \(\phi^I_{(J)}(u_*)={\delta^I}_J\) without loss of generality.
We can construct a matrix $H$ as follows:
\begin{equation}
	{H(\omega,k_{\perp},u)^I}_J = \phi^I_{(J)}(u) ~~\text{with}~~
	 {H(u_*)^I}_J = {\delta^I}_J ,
\end{equation}
which provides $\bar{\Phi}_{\text{reg}}(u) = H(u)\bar{\varphi}$. 
Since $\bar{\Phi}_{\text{reg}}(0)=0$ is required at the AdS boundary,
the mode frequency $\omega$ can be determined by the equation 
\begin{equation}
	H_{0}(\omega, k_{\perp}) \equiv
	\lim_{u\to0} \det\left[
		{H(\omega,k_{\perp}, u)^I}_J
	\right] = 0,
	\label{eq:det=0}
\end{equation}
and the mode functions are determined by solving 
$H(0)\bar{\varphi} = 0$, meaning that $\bar{\varphi}$ is an
eigenvector for $H(0)$ with eigenvalue $0$.
Meanwhile, if $\det H(0) \neq 0$, we write an ingoing-wave solution as follows:
\begin{equation}
	\bar{\Phi}^I_{\text{reg}}(u) = {F^{I}}_J(\omega,k_{\perp}, u) \bar{\Phi}^J_{\text{reg}}(0),
	\label{eq:bulk-to-boundary}
\end{equation}
where matrix \(F\) is a bulk-to-boundary propagator related to a matrix \(H\):
\begin{equation}
	F(u) = H(u) H(0)^{-1}.
\end{equation}

In the sense of the field theory, the frequencies of the modes correspond to the poles of Green's functions.
The retarded Green's function in the NESS can be computed by%
\footnote{
	Precisely, we must consider another term from a counterterm of the action \cite{Karch:2005ms}.
	The counterterm is needed to regulate the action at the AdS boundary.
	However, we can ignore it for our purpose to find QNMs.
}
\begin{equation}
	G^{R}(\omega,k_{\perp}) =
	\lim_{u\to 0} \left[
		F_{-k}^{T} \bar{A}_{k} F_{k}' + F_{-k}^{T} \bar{B}_{k}^{\dagger} F_{k}
	\right].
	\label{eq:NESS_Green}
\end{equation}
As mentioned above, \(F\) is ill-defined when eq.~(\ref{eq:det=0}) is satisfied,
implying that the frequencies satisfying eq.~(\ref{eq:det=0}) equal the poles of (\ref{eq:NESS_Green}).
Studying the Green's function in the NESS is also interesting, but we focus on the poles of the Green's function in this study. 

\begin{figure}[htbp]
	\centering
	\includegraphics[width=0.6\linewidth]{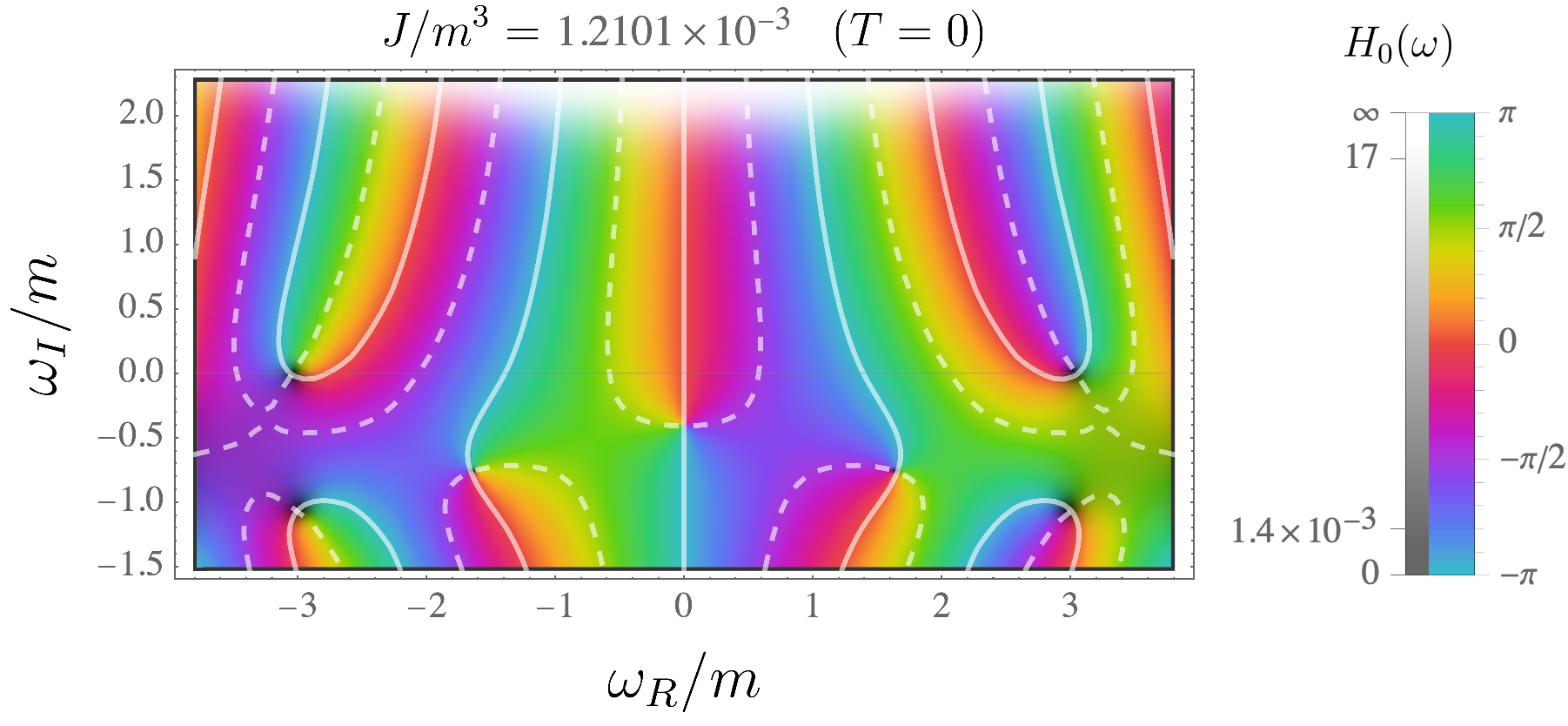}
 \caption{
		Example of complex function \(H_0(\omega)\) in complex
 $\omega$ plane: \(\omega = \omega_R + i \omega_I\).
		The white solid lines show \(\mathrm{Im}H_0(\omega)= 0\).
		The white dotted lines show \(\mathrm{Re}H_0(\omega) = 0\).
		Points where the solid and dotted lines cross correspond to the zeros or the poles of \(H_0(\omega)\).
	}
	\label{fig:cmaps_ex}
\end{figure}
Figure \ref{fig:cmaps_ex} shows the argument of 
\(H_0(\omega, k_{\perp} = 0)\)
defined by eq.~(\ref{eq:det=0}) as a complex function 
in the complex $\omega$ plane for a background solution.
From this figure, one can read mode frequencies, i.e., poles of the Green's function.
According to the determinant method, the poles are located where  $H_0(\omega) = 0$, corresponding to the intersections of the lines of $\mathrm{Re}H_0(\omega)= 0$ and $\mathrm{Im}H_0(\omega) = 0$ in figure \ref{fig:cmaps_ex}.

\section{Dynamical stability in the conducting phase}\label{sec:Results}
In this section, we describe the dynamical stability of our system in the
conducting phase.
From section~\ref{sec:Background}, multiple solutions with various $J$ emerge for a given electric field, $E$, or temperature, $T$, in the conducting phase.
The stability of our system can be investigated by solving the perturbation equations.
We find instability associated with the multivalued property of the $E$--$J$ curves in the system.
Further, there are static solutions of the perturbation fields with a specific spatial scale, implying that the system also has an inhomogeneous background solution.

\subsection{Stability for homogeneous perturbations and multivaluedness}
\subsubsection{Zero temperature}
\begin{figure}[htbp]
	\centering
	\includegraphics[width=0.49\linewidth]{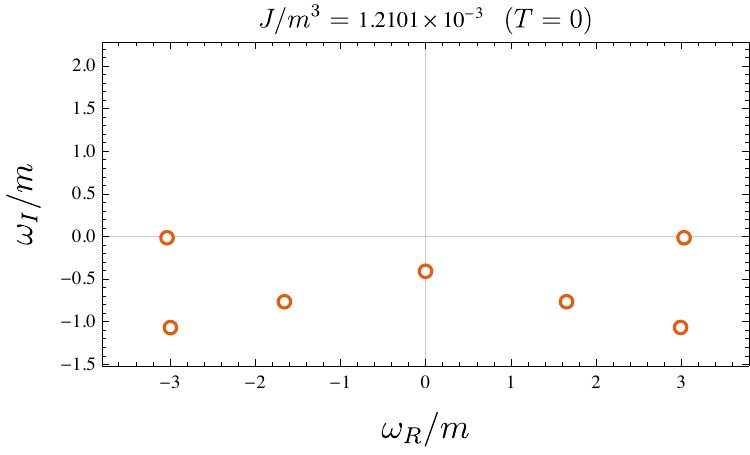}
	\includegraphics[width=0.49\linewidth]{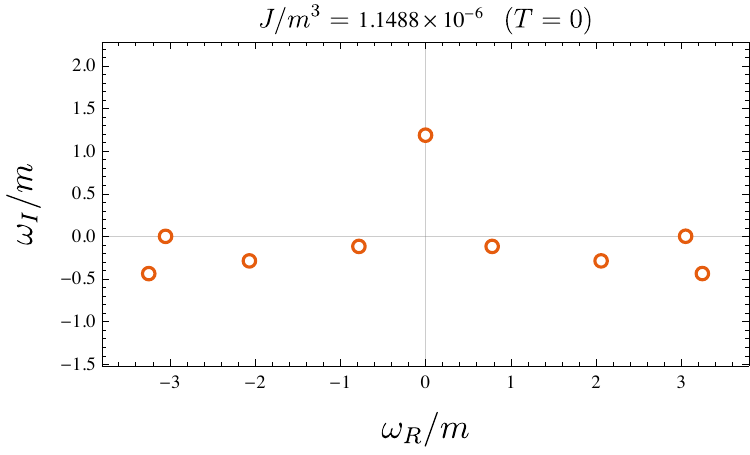}
	\caption{
		Frequencies of the QNMs on the complex plane of 
		\(\omega = \omega_R + i \omega_I\) for some \(J/m^3\);
		the left panel shows $J/m^3 = 1.2101\times 10^{-3}$ on the upper branch 
		and the right panel shows $J/m^3 = 1.1488\times 10^{-6}$ on the lower branch.
	}
	\label{fig:cmaps}
\end{figure}

\begin{figure}[htbp]
	\centering
	\includegraphics[width=0.49\linewidth]{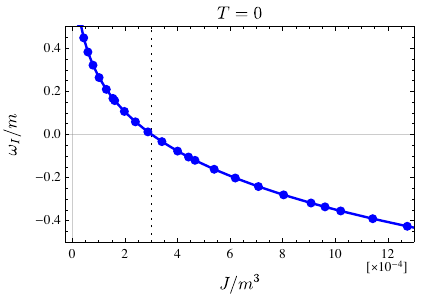}
	\caption{
		\(\omega_I/m\) of the pure-imaginary mode as a
		function of \(J/m^3\) at \(T=0\).
		The points on the curve denote numerical data.
		The vertical dotted line indicates
		\(J^{(1)}_{\text{turn}}/m^3 = 3.0031 \times 10^{-4}\) at the first turning point.
	}
	\label{fig:oi-J_first}
\end{figure}

First, we show the behavior of the mode frequencies with respect to homogeneous perturbations of \(\delta A_x\) and \(\delta \theta\) at zero temperature with a finite current.
As shown in the left panel of figure \ref{fig:zeroT}, two branches in
the background solutions appear for $E \ge E^{(1)}_{\rm turn}$;
for a given electric field, 
the upper (lower) branch has a higher (lower) electric current 
than $J^{(1)}_{\text{turn}}/m^3 = 3.0031 \times 10^{-4}$.
In figure \ref{fig:cmaps} we show some frequencies of the QNMs, i.e., the poles of the retarded Green's functions, 
in the complex $\omega$ plane.
The left and right panels correspond
to background solutions with $J/m^3 = 1.2101\times 10^{-3}$ on the
upper branch and a background 
solution with $J/m^3 = 1.1488\times 10^{-6}$ on the lower branch, respectively.
We find that, in the left panel (the upper branch), no pole locates in the upper-half plane of $\omega$, whereas, in the right panel (the lower branch), a pole on the imaginary axis, 
$\omega/m \simeq 1.2 i$, has moved to the upper-half plane of $\omega$.
In fact, it turns out that this pure-imaginary mode is
the first going across the real axis at the turning point
as $J$ decreases from the upper to lower branch.%
\footnote{
	From the left panel of figure \ref{fig:cmaps}, one can find poles at \(\omega/m \approx \pm 3.0\).
	In this figure, these poles have very small negative imaginary part. 
	We have checked these poles also locate on the lower-half plane when the background solution is the upper branch solution.
}
Figure \ref{fig:oi-J_first} shows \(\omega_I/m\) of the pure-imaginary
mode as a function of \(J/m^3\) around the first turning point.
Thus, we conclude that the upper branch solutions with higher $J$ are stable,
whereas the lower branch ones with lower $J$ become unstable against the
corresponding linear perturbations.

\begin{figure}[htbp]
	\centering
	\includegraphics[width=0.49\linewidth]{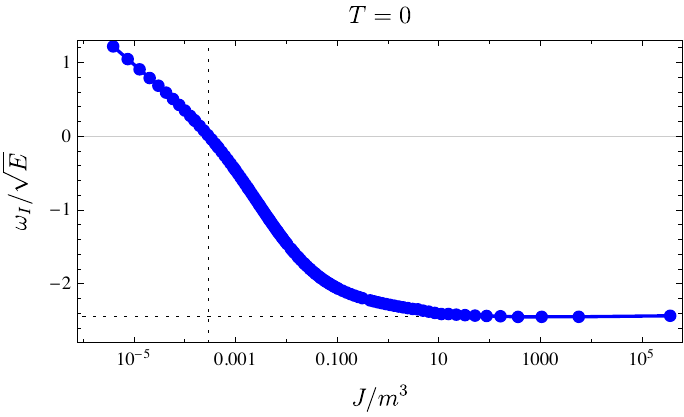}
	\caption{
		Asymptotic behavior of the pure-imaginary mode for large \(J/m^3\) at \(T=0\).
		The horizontal axis is shown in log scale. 
		The vertical dotted line shows \(J/m^3 = J^{(1)}_{\text{turn}}/m^3\).
		The horizontal dotted line shows 
		\(\omega_I/\sqrt{E} = - \sqrt{6}\), which is the analytic value in the
		massless limit $m=0$.
	}
	\label{fig:oi-J_large}
\end{figure}
In addition, we observe the asymptotic behavior of the pure-imaginary mode
for large \(J/m^3\).
Figure \ref{fig:oi-J_large} shows \(\omega_I/\sqrt{E}\) as a function of \(J/m^3\).
We find that this mode locates on the imaginary axis for any \(J/m^3\),
and \(\omega_I/\sqrt{E}\) seems to converge to \(-\sqrt{6}\) in the limit of \(J/m^3 \to \infty\).
In the massless quark case, i.e., $m=0$, the current is
analytically given by $J=E^{3/2}$.
This asymptotic value equals \(-\kappa/\sqrt{E}\) for \(m=0\), where 
\(\kappa\) denotes the surface gravity on the effective horizon defined in eq.~(\ref{eq:surface_gravity}).%
\footnote{
	An equation of motion for the perturbation of the embedding function in the case of massless quark and applying \(E\) has been derived in ref.\,\cite{Albash:2007bq}.
}

\begin{figure}[htbp]
	\centering
	\includegraphics[width=0.49\linewidth]{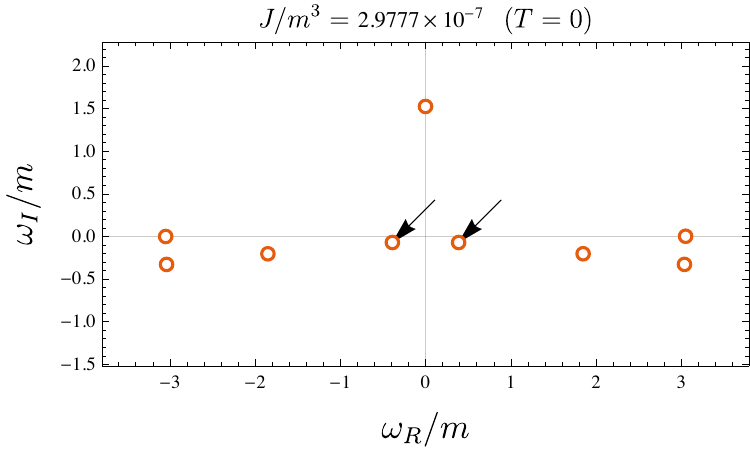}
	\includegraphics[width=0.49\linewidth]{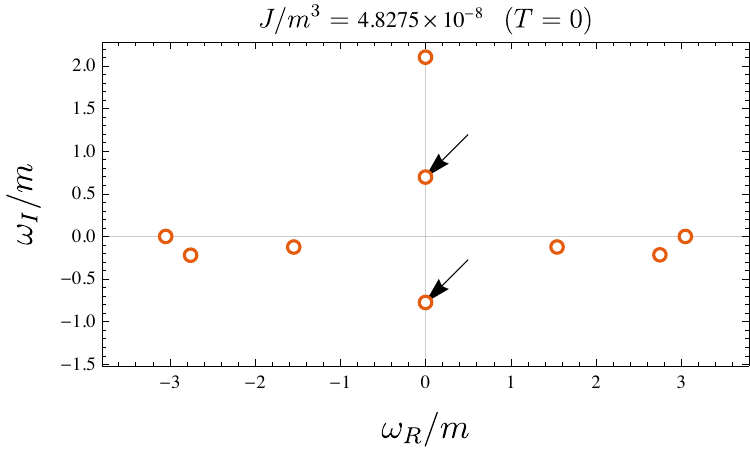}
	\caption{
		Frequencies of the QNMs on the complex plane of 
		\(\omega = \omega_R + i \omega_I\) around the second turning point of
		$E$--$J$ curve.
		In each panel the two poles pointed by the arrows involve the second turning point. 
	}
	\label{fig:cmaps_second}
\end{figure}

Then, we also examine the behavior of poles around the second turning
point of the $E$--$J$ curve at \(T=0\), shown in the right panel of figure \ref{fig:zeroT}.
The second turning point is located at \(J^{(2)}_{\text{turn}}/m^3 = 1.9178 \times 10^{-7}\).
We show 
mode frequencies on the complex $\omega$ plane for 
$J/m^3 = 2.9777\times 10^{-7}$ and $4.8275\times 10^{-8}$
in figure \ref{fig:cmaps_second}.
We will refer to a set of solutions within \(1.9178\times 10^{-7} < J/m^3 < 3.0031\times 10^{-4}\) as the (second) upper branch, and a set of solutions with \(J/m^3 < 1.9178\times 10^{-7}\) as the (second) lower branch.
The left and right panels of figure \ref{fig:cmaps_second} correspond to solutions of the (second) upper and lower branches in the right panel of figure \ref{fig:zeroT}, respectively.
Notably, there exists one more branch with a higher $J$, which is
represented in the left panel of figure \ref{fig:zeroT}.
Because the $E$--$J$ curve has already been multivalued for a given
$E/m^2$ in \(J/m^3\), both branches extending from the second turning point have 
an unstable mode on the imaginary axis, as seen previously.
\begin{figure}
	\centering
	\includegraphics[width=0.9\linewidth]{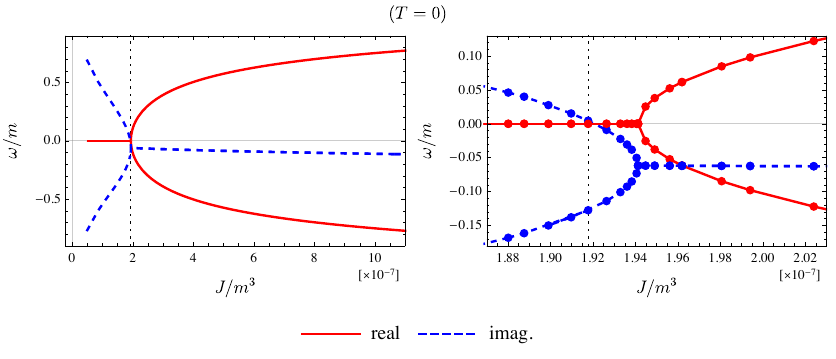}
	\caption{
		(left)
		\(\omega/m\) of the two poles pointed in figure \ref{fig:cmaps_second} as functions of \(J/m^3\).
		The red solid line and the blue dashed line represent
 the real and imaginary part of \(\omega\), respectively.
		The dotted vertical line shows \(J/m^3\) of the second turning point: \(J^{(2)}_{\text{turn}}/m^3 = 1.9178 \times 10^{-7}\).
		(right)
		An enlarged view of the left panel around \(J^{(2)}_{\text{turn}}/m^3 = 1.9178 \times 10^{-7}\).
	}
	\label{fig:Jgap}
\end{figure}
In the left panel ($J/m^3 = 2.9777\times 10^{-7}$), 
we find two poles near the origin of \(\omega\) in the lower-half plane;
so, these are stable modes.
Meanwhile, in the right panel ($J/m^3 = 4.8275\times 10^{-8}$),
the corresponding poles become two pure-imaginary modes newly added on the
imaginary axis, one of which is located in the upper-half plane and is unstable.
To illustrate the behavior of the two poles, we show \(\omega\) of these poles as functions of \(J/m^{3}\) in figure \ref{fig:Jgap}.
As $J$ decreases, these two poles first approach each other in the
lower-half plane to meet on the imaginary axis and then move apart on the imaginary axis.
We find that the second turning point corresponds to \(\omega = 0\) for
one of these poles.

At the first and second turning points of the $E$--$J$ curve, we
observed a pole with $\omega=0$.
It seems to be quite natural for the following reasons.
At a turning point of the $E$--$J$ curve, 
the background solution will bifurcate into two branches with different
$J$ for a given $E$, implying a static perturbation to an infinitesimally change of $J$ for a
fixed $E$.
Moreover, 
the results agree with an expectation that \(\pdv*{J}{E}\) is related to the retarded Green's function by the Green-Kubo formula, even in the NESS.
The divergence of \(\pdv*{J}{E}\) corresponds to the existence of a pole at zero.
We expect that a pole with $\omega=0$ corresponding to a static
perturbation will emerge at every other turning point.

\subsubsection{Finite temperature}


We now proceed to study the dynamical stability at a finite temperature.
We focus only on a pure-imaginary mode because this mode will govern the stability of the system.

\begin{figure}[htbp]
	\centering
	\includegraphics[width=0.98\linewidth]{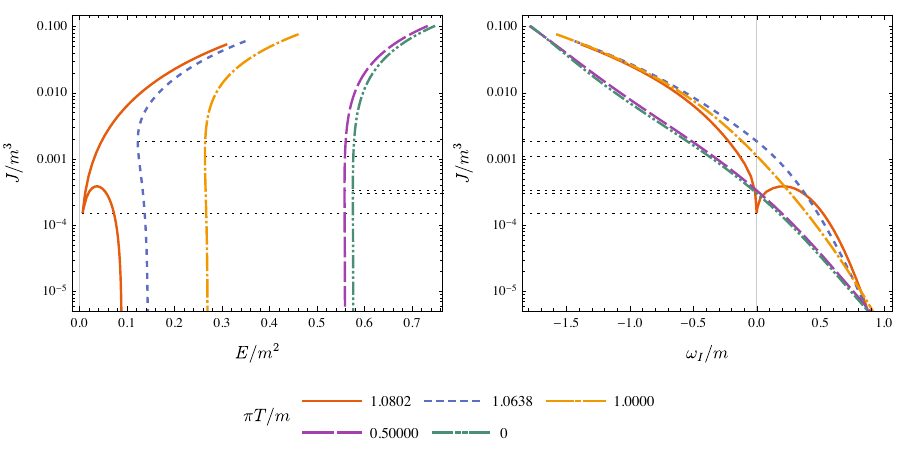}
	\caption{
		(left) $E$--$J$ curves at various temperatures.
		(right) Current \(J/m^{3}\) as a function of the imaginary part of mode frequency \(\omega_I/m\) for the pure-imaginary mode.
		The horizontal dotted lines in the both panels show the first turning points of $E$--$J$ curves at each temperatures, and these points completely match to points where \(\omega_{I}\) goes across zero.
	}
	\label{fig:oi-J}
\end{figure}
The relation between the instability and the multivalued property of the $E$--$J$ curves is more clearly illustrated in figure \ref{fig:oi-J}.
It shows the $E$--$J$ curves and the relation between
\(J\) and the imaginary part of mode frequency \(\omega_I\) for the
pure-imaginary mode at various temperatures.
In the vicinity of the critical electric field or small-current region, \(J\) becomes a multivalued function of \(E\).
In the multi-valued region, there are mostly two branches of background
solutions: 
the upper branches with higher $J$ and lower branches with lower $J$ for the
given $E$ and $T$.
We observe that \(\omega_{I}\) is negative and positive throughout the upper and lower branches, respectively.
The zero-points of \(\omega_{I}\) completely match the line of the first turning points in the $E$--$J$ curves (figure \ref{fig:oi-J}).
These results imply that the lower branches are unstable against 
homogeneous 
perturbations of the current or quark condensate.

\subsection{Inhomogeneous perturbations: spatial instability}

\begin{figure}[htbp]
	\centering
	\includegraphics[width=0.65\linewidth]{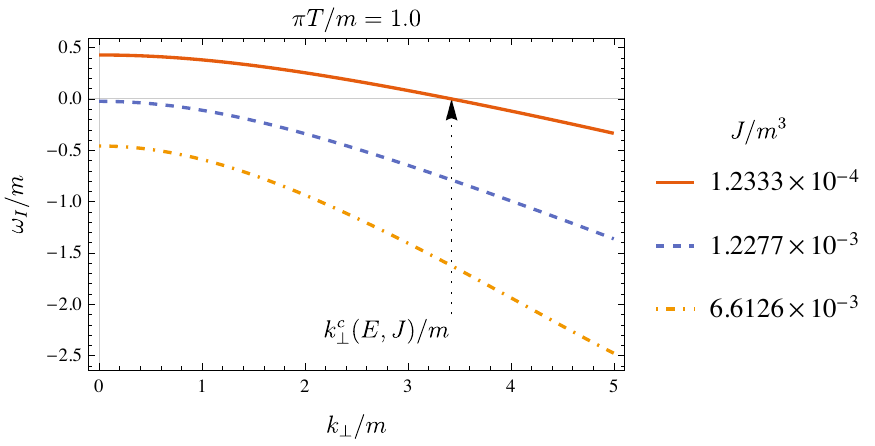}
	\caption{
		\(\omega_I/m\) as functions of \(k_{\perp}/m\) for several \(J/m^3\) at \(\pi T/m = 1.0000\).
		For \(J/m^3 = 1.2333\times 10^{-4}\), \(\omega_I/m\) goes across zero at \(k_{\perp}/m \approx 3.4\).
		Such a critical wavenumber $k_{\perp}^c$ depends on \(E\) and \(J\) of the background solutions.
	}
	\label{fig:oi-k}
\end{figure}
Now, we consider inhomogeneous perturbations with finite wavenumber
$k_{\perp}$ perpendicular to the electric field and current.
As mentioned above, in the multivalued region near the critical
electric fields the background solutions of the lower-$J$ branch have  
instability, resulting from the pure-imaginary mode for the homogeneous
perturbations.
Figure \ref{fig:oi-k} shows the imaginary part of frequency \(\omega_I\)
for the
pure-imaginary mode as functions of \(k_{\perp}\) for several points on the $E$--$J$ curve at \(\pi T / m = 1.0000\).
Because $k_{\perp} = 0$ means the homogeneous perturbations, the electric
currents 
$J/m^3=1.2333\times 10^{-4}$, $1.2277\times 10^{-3}$, and 
$6.6126\times 10^{-3}$ 
represent an unstable solution on the lower branch, marginal one on the
turning point, and stable one on the upper branch, respectively. 
We observe \(\omega_I\) decreases as \(k_{\perp}\) increases for each $J$.
In particular, \(\omega_I\) can go across zero at a specific wavenumber
\(k_{\perp} = k_{\perp}^c\) even if \(\omega_I\) is positive at
\(k_{\perp}=0\), i.e., the homogeneous perturbation is 
a growing mode, 
meaning that our system becomes unstable against perturbations below a
critical wavenumber, and this instability is a long-wavelength instability.
We can find such \(k_{\perp}^c\) for the background solutions on the lower branch which has positive \(\omega_I\) when \(k_{\perp}=0\).
Because the perturbative solution can be both static and inhomogeneous with a
specific length scale, \(1/k_{\perp}^c\),
the existence of \(k_{\perp}^c\) implies that the system also has
inhomogeneous nonlinear solutions such that the current density is
spatially modulated (figure \ref{fig:intro}).

\begin{figure}[htbp]
	\centering
	\includegraphics[width=0.8\linewidth]{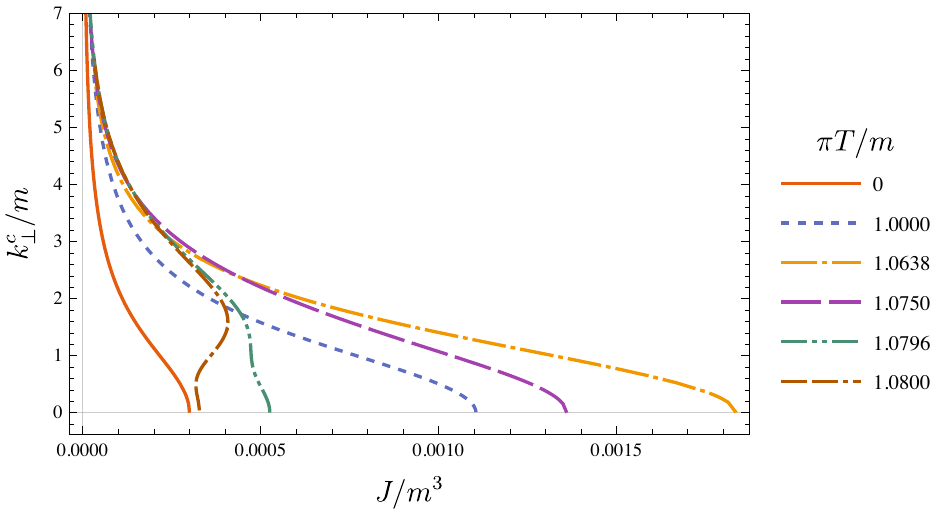}
	\caption{
		\(k_{\perp}^c/m\) vs \(J/m^{3}\) at various temperature.
			For \(\pi T/m \geq 1.0796\), \(k_{\perp}^c/m\) becomes multi-valued.
	}
	\label{fig:k-J}
\end{figure}
\begin{figure}[htbp]
	\centering
	\includegraphics[width=0.7\linewidth]{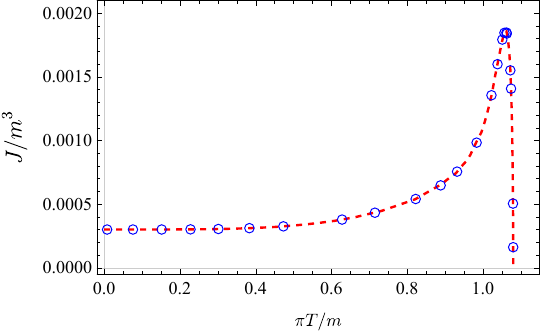}
	\caption{
 The relation between the value of $J$ at $k^{c}_{\perp}=0$ and
 temperature $T$ (the blue open circles).
That corresponds to endpoints of each curve at $k^{c}_{\perp}=0$ in
figure \ref{fig:k-J}.
The red dashed line denotes the first turning point on each $E$--$J$
 curve, which is identical to that shown in the left panel of figure
 \ref{fig:ETJ}. 
The plot shows that the turning points, 
at which the zero-modes with $\omega = k_\perp =0$ appear,
have a non-monotonic dependence on the temperature.
	}
	\label{fig:turnplot}
\end{figure}
\begin{figure}[htbp]
	\centering
	\includegraphics[width=0.8\linewidth]{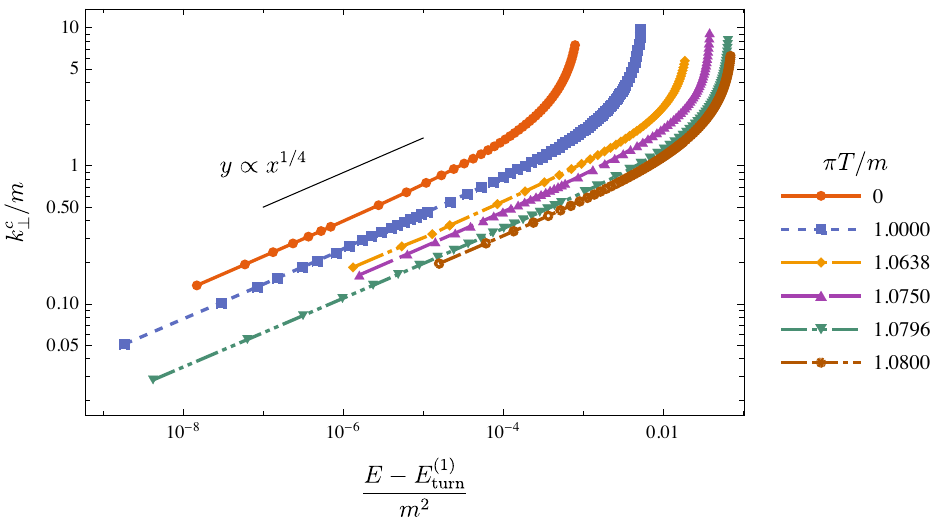}
	\caption{
		\(k_{\perp}^c/m\) as functions of \((E-E^{(1)}_{\text{turn}})/m^2\) at various temperatures.
		Both the vertical and horizontal axes are shown in log scale.
		The black line shows \(y\propto x^{1/4}\) for reference.
	}
	\label{fig:kc-E}
\end{figure}
Figure \ref{fig:k-J} shows relations between \(k_{\perp}^c\) and \(J\) at various temperatures.
For \(\pi T/m \geq 1.0796\), \(k_{\perp}^c\) becomes multivalued to \(J/m^3\) because there are two unstable solutions for given \(J/m^3\) in the $E$--$J$ curves.
The endpoints where \(k_{\perp}^c=0\) correspond to the first turning points for each temperature (figure \ref{fig:turnplot}).
In the vicinity of \(J=0\), \(k_{\perp}^c\) blows up,
meaning that this instability will extend to any short length-scale toward the critical embedding solutions.

We also show relations between \(k_{\perp}^c\) and \(E\) in figure \ref{fig:kc-E},
where we have plotted \(k_{\perp}^c/m\) as functions of \((E-E^{(1)}_{\text{turn}})/m^2\) in log-log scale at various temperatures.
From this figure, \(k_{\perp}^c/m\) behaves as 
a power 
function of \((E-E^{(1)}_{\text{turn}})/m^2\) for each temperature.
We also read the power-law index in the vicinity of \(E = E^{(1)}_{\text{turn}}\) by a linear fitting.
If we write the relation in the vicinity of \(E = E^{(1)}_{\text{turn}}\) as
\begin{equation}
	\frac{k_{\perp}^c}{m}
	\approx
	c
	\left(
		\frac{E - E^{(1)}_{\text{turn}}}{m^2}
	\right)^{p},
\end{equation}
where \(c\) and \(p\) are constant,
the exponents \(p\) at various temperatures are estimated as follows:
\begin{equation}
	\begin{array}{c | c c c}
	\pi T/m & 0 & 1.0000 & 1.0796\\ \hline
	p & 0.260 & 0.257 & 0.251
	\end{array}
\end{equation}
These values are estimated from the data around small but finite \((E-E^{(1)}_{\text{turn}})/m^2\).
We expect that the true exponent will be given by \(p=1/4\) for any temperature.


\section{Conclusion and discussion}\label{sec:conclusion_and_discussion}
In this article, we study the D3-D7 model under an electric field. We focus on the NESS system with a constant current in the dual field theory. Moreover, we perform a linear stability analysis of the background states.

In section \ref{sec:Background}, we present the phase diagram of the conducting/insulating phases and the behaviors of the current density, quark condensate, and effective temperature as functions of \(E\) and \(T\).
Figure \ref{fig:ETJ} shows the multivalued $J$ with respect to $E$ for various $T$.
On decreasing \(E\) or \(J\), we find the first turning point in the $E$--$J$ curves.
At smaller \(J\), we also find the second turning point (figure~\ref{fig:zeroT}).
At \(E=0\), a similar multivaluedness of the parameters in the D3-D7 model has been found in refs.~\cite{Mateos:2006nu,Mateos:2007vn,Frolov:2006tc}.
We expect that the turning points will repeatedly appear in the $E$--$J$ curves until reaching \(J=0\).
This behavior of the probe brane system in the black hole geometry may be understood as the discrete scale invariance \cite{Sornette_1998}.
Similar behaviors could be observed experimentally in an insulator in the vicinity of the electric breakdown.

To study the linear stability of the system, we consider the current perturbation of the background solution in the multivalued region.
On the gravity side, this corresponds to the perturbation of the worldvolume gauge field parallel to the external electric field being coupled with the perturbation of the brane embedding. 
By solving the equations for the coupled perturbations and computing the QNMs, we find the dynamical instability of the NESS system in the multivalued region.
At any temperature, the upper and lower branches of the $E$--$J$ curves are, respectively, stable and unstable against the homogeneous perturbations.

The authors of \cite{Kaminski:2009ce} showed similar observations
in the D3-D7 model without applying $E$, i.e., an equilibrium system.
In equilibrium, the on-shell action of the probe brane is regarded as a thermodynamic potential of the boundary theory \cite{Mateos:2006nu,Mateos:2007vn}.
It has also been shown that solutions corresponding to the unstable branch of the thermodynamic potential are dynamically unstable from the QNM analysis.
On the unstable branch, the quark condensate is multivalued for a given quark mass.
Their unstable solutions with vanishing density correspond to our unstable solutions in the limit of \(E=0\).
So far, in this study, we focus on the dynamical stability of the system with respect to small perturbations.
It is also intriguing to consider about thermodynamic stability of the NESS.
If we can define a meaningful ``thermodynamic potential,'' even for the NESS system, it might show a swallow-tail structure for $E$ and determine the first-order transition point.
Our results indicate the dynamical instability of the NESS system, but the thermodynamic stability is uncertain.

By considering the finite wavenumber for the perturbations around an unstable branch,
we find the critical wavenumber with which the perturbations
do not depend on time, meaning the background solution with the static perturbation, which has the specific scale of the critical wavelength, can be realized.
It also implies that there are inhomogeneous background states as new branches around the unstable solutions.
Current filaments can be realized in conductors that show negative differential conductivity \cite{Schoell:2001}.
We expect that our observations indicate the holographic realization of the current filaments.
Notably, the perturbation becomes unstable in the long-wavelength region (figure \ref{fig:oi-k}).
Because the unstable modes do not appear in the system with the finite size $L \lesssim 1/k_{c}$, it is expected that a homogeneous state also becomes stable.
In addition, we remark that the direction of the inhomogeneity is symmetric under the rotation in \(y\)-\(z\) plane because the system is isotropic in this plane. Assuming that the phase transition from a homogeneous state to an inhomogeneous state occurs, the rotational symmetry will be spontaneously broken, or explicitly broken by perturbing along a specific direction.

Several open questions should be explored in the future.
First, the relation between dynamical and thermodynamic stabilities in the NESS system should be examined.
Holographically, a naive candidate for
the ``thermodynamic potential'' in the NESS system might be the probe brane action as in equilibrium.%
\footnote{
	For example, the authors of
	\cite{Matsumoto:2018ukk,Imaizumi:2019byu} studied the several
	critical exponents by using the on-shell DBI action as the
	thermodynamic potential under the same setup of the NESS
	system. A similar definition was also given in \cite{Banerjee:2015cvy}.
}
It would be interesting to compare our results with the ``thermodynamic potential'' in the same parameter region.

Another direction is to find inhomogeneous nonlinear solutions, which are indicated by our linear stability analysis, apart from the linear perturbation analysis.
To do this, one must solve the nonlinear partial differential equations as the equations of motion for the fields in the D3-D7 model. 
Thus, one can investigate the stability of the inhomogeneous solutions.

It would be crucial to understand the origin of the dynamical instability in our setup.
The authors of \cite{Nakamura:2009tf} found a filamentary instability by considering a five-dimensional Maxwell theory with the Chern-Simons (CS) term in AdS\(_5\times \mathrm{S}^5\) spacetime.
A similar instability has been observed in the D3-D7' model which duals to the gauge field theory with quark-like particles in \((2+1)\)-dimensions by the authors of  \cite{Bergman:2011rf}.
In both cases, the considered system becomes dynamically unstable at a specific finite range of the wavenumber, unlike our case.
This type of filamentary instability originated from the CS term in the gravity model action.
In our system, because the CS term does not affect the perturbations in the linear order, the filamentary instability must have another origin. This is left for future work.

\begin{acknowledgments}
The authors are grateful to Shin Nakamura and Yuichi Fukazawa for the fruitful discussion.
S.~I.~is supported by National Natural Science Foundation of China with Grant No.~12147158.
M.~M.~is supported by National Natural Science Foundation of China with Grant No.~12047538.
S.~K.~is supported by JSPS KAKENHI Grant Numbers~JP16K17704, JP21H05189.
The authors thank RIKEN iTHEMS NEW working group for fruitful discussions.
The authors would like to thank Enago (www.enago.jp) for the English
 language review.
\end{acknowledgments}

\appendix

\section{Numerical details}
In this study, we have mainly used a shooting method for computing the QNMs.
To solve eq.~(\ref{eq:linear_eom_kspace}) numerically, we need to
introduce a small cutoff around the singularities of the differential equations at \(u = u_*\) and \(u=0\).
Thus, the integral region becomes \(\varepsilon_{\text{UV}} \leq u \leq u_* -\varepsilon_{\text{IR}}\) where \(\varepsilon_{\text{UV},\text{IR}}\) are numerical cutoffs.
With the ingoing-wave boundary condition, we can obtain a series solution for the perturbation fields around \(u=u_*\).
We can solve eq.~(\ref{eq:linear_eom_kspace}) numerically by imposing a condition at \(u=u_* - \varepsilon_{\text{IR}}\) which claims that the numerical solution equals the series solution at this point.
In our computation, we used \(\varepsilon_{\text{IR}}/u_* = 0.0001,\) \(\varepsilon_{\text{UV}}/u_* = 0.00001\) and the series solution up to the first order of \((u-u_*)\).

\begin{figure}[htbp]
	\centering
	\includegraphics[width=0.7\linewidth]{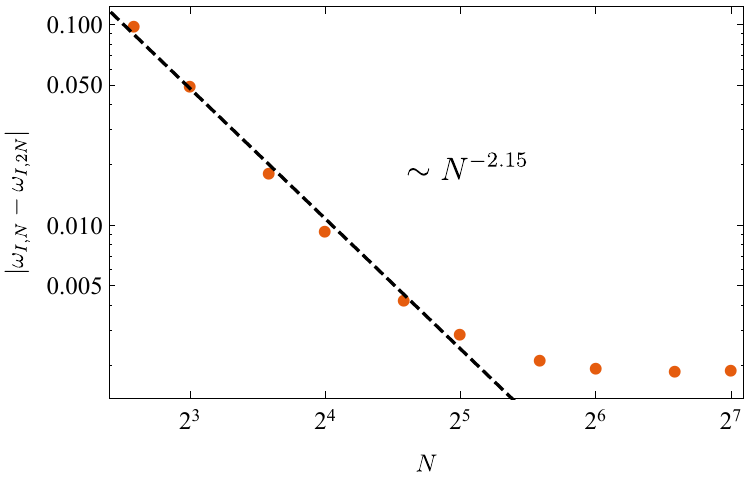}
	\caption{
	The convergence behavior of $\left| \omega_{I,N}-\omega_{I,2N} \right|$ with respect to the numbers of the grid points. Here, we use $E/m^{2}=1.0627$ and $J/m^{3}=0.39901$.
	}
	\label{fig:convergence}
\end{figure}

The authors of  \cite{Kaminski:2009ce} pointed out that the shooting method could be difficult for computing QNMs with large \(-\omega_{I}\).
Because we focus on the QNMs with small \(|\omega_{I}|\), we expect to avoid the difficulty.
To ensure this, we attempted to use a relaxation method for computing the QNMs.
In the relaxation method, we discretized the equations of motion with the second-order finite differences on the $N$ grid points. Using an appropriate initial configuration, we obtained the numerical solutions after iteration in the Newton-Raphson relaxation scheme. We stopped the iteration when the root mean squared error with respect to the equations of motion was less than $10^{-10}$. We also checked the convergence of the variables with respect to the number of grid points. The discretization error in the second-order differences was expected to be given by the square of the grid size. Thus, if we double the grid points to $2N$, the discretization error should be reduced to a quarter of that in $N$ grid points.
Figure \ref{fig:convergence} shows the typical convergence behavior of $\omega_{I}$ with respect to the number of grid points $N$. We define the numerical error as $\left| \omega_{I,N}-\omega_{I,2N} \right|$, where $\omega_{I,N}$ is the numerical value of the quasinormal frequency with $N$ grid points.
Figure \ref{fig:convergence} indicates that the numerical error converges as the number of  grid points $N$ increases.
For $N\lesssim2^{5}$, the numerical error converges as $N^{-2.15}$, which roughly agrees with that expected from the discretization error, namely $N^{-2}$. Moreover, because the convergence is not improved for $N\gtrsim2^{6}$, the improvement of the numerical accuracy is also unexpected for more grid points.
Notably, the background solutions $\theta(u)$ and $h(u)$, which are also numerically obtained, may affect the numerical error of the relaxation method. The convergent behavior of the quantity with respect to the grid points implies that the numerical scheme works well for our problem.

\begin{figure}[htbp]
	\centering
	\includegraphics[width=0.99\linewidth]{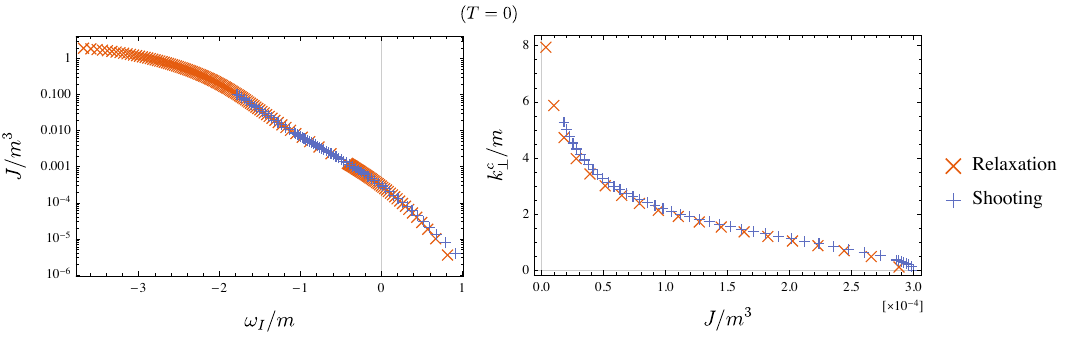}
	\caption{
		(left) \(J/m^3\) vs \(\omega_I/m\),
		(right) \(k_c^{\perp}/m\) vs \(J/m^3\) at \(T=0\) computed by the shooting method and the relaxation method.
	}
	\label{fig:oi,kc-J_comp}
\end{figure}
We show the comparison of the shooting and relaxation method and the relaxation in figure \ref{fig:oi,kc-J_comp}.
We plot \(J/m^3\) as a function of \(\omega_I/m\) at \(T=0\) using two different methods.
The data of the shooting method are the same as in figure \ref{fig:oi-J} and figure \ref{fig:k-J}.
The results computed from the two methods almost agree with each others.
Consequently, our results in the shooting method do not encounter the difficulty discussed in ref.~\cite{Kaminski:2009ce}.

\section{Ingoing-wave boundary condition}\label{appendix:ingoing}
We now discuss the ingoing-wave  boundary condition from the effective action of the perturbations in the position space.
It is given by
\begin{equation}
	S^{(2)} = -\frac{\mathcal{N}}{2}\int \dd[4]{x}\dd{u} R(u)\Big[
		\partial_{\alpha} \tilde{\Phi}^T \tilde{A}^{\alpha\beta} \partial_{\beta} \tilde{\Phi}
  		+ 2 \tilde{\Phi}^T \tilde{B}^{\alpha} \partial_{\alpha} \tilde{\Phi}
  		+ \tilde{\Phi}^T \tilde{C} \tilde{\Phi}
	\Big],
	\label{eq:effective_action_Appendix}
\end{equation}
with coefficient matrices:
\begin{subequations}
\begin{gather}
	\tilde{A}^{\alpha\beta}
	=
	\gamma^{\alpha\beta}P(u) +
	2 \delta^{[\alpha}_{t} \delta^{\beta]}_u Q(u),\\
	\tilde{B}^{\alpha} = 3 \tan\theta
	\begin{bmatrix}
		- M^{\alpha u} \theta'(u) & M^{\alpha x}\\
		0 & 0
	\end{bmatrix},~~~
	\tilde{C} = 
	- 3 (1 - 2\tan \theta)
	\begin{bmatrix}
		1 & 0\\
		0 & 0
	\end{bmatrix},
\end{gather}
\label{eq:coeffs_position}
\end{subequations}
where
\begin{subequations}
\begin{gather}
	P(u) \equiv
	\begin{bmatrix}
		\Xi(u) & -M^{xu} \theta'(u)\\
		-M^{xu} \theta'(u) & \gamma^{xx}
	\end{bmatrix},\\
	Q(u) \equiv
	(M^{tx} \gamma^{uu} + \gamma^{tu} M^{xu})\theta'(u)
	\begin{bmatrix}
		0 & 1\\
		-1 & 0
	\end{bmatrix}.
\end{gather}
\end{subequations}
\(\Xi(u)\) is defined by eq.~(\ref{eq:Xi_Omega_definition}).
\(P\) is a symmetric matrix, and \(Q\) is an antisymmetric matrix.
\(\tilde{A}^{\alpha\beta}\) satisfies \((\tilde{A}^{\alpha\beta})^T = \tilde{A}^{\beta\alpha}\) by definition.
\(P\) is diagonalized as
\begin{equation}
	U^T P U =
	\begin{bmatrix}
		\frac{\gamma^{xx} + \Xi(u) - \Sigma(u)}{2} & 0\\
		0 & \frac{\gamma^{xx} + \Xi(u) + \Sigma(u)}{2}
	\end{bmatrix}
	\equiv
	\begin{bmatrix}
		p_1 & 0\\
		0& p_2
	\end{bmatrix},
\end{equation}
with an orthogonal matrix
\begin{equation}
	U(u) =
	\frac{1}{\sqrt{2}}
	\begin{bmatrix}
		\sqrt{
			\frac{\gamma^{xx} - \Xi(u) + \Sigma(u)}{\Sigma(u)}
		}
		&
		-\sqrt{
			\frac{-\gamma^{xx}+\Xi(u)+\Sigma(u)}{\Sigma(u)}
		}\\
		\sqrt{
			\frac{- \gamma^{xx} + \Xi(u) + \Sigma(u)}{\Sigma(u)}
		} &
		\sqrt{
			\frac{\gamma^{xx} - \Xi(u) + \Sigma(u)}{\Sigma(u)}
		}
	\end{bmatrix}.
\end{equation}
In the above equations, we have defined
\begin{equation}
	\Sigma(u) \equiv \sqrt{(\gamma^{xx}-\Xi(u))^2 + 4 (M^{xu})^2\theta'(u)^2}.
\end{equation}
We write \(\hat{P} \equiv U^T P U\).
\(p_1\) and \(p_2\) are real because \(P\) is a real symmetric matrix.
\(U^T Q U = Q\) under the transformation because it is a \(2\times 2\) totally antisymmetric matrix.

Because we are interested in the behavior of the fields in the vicinity of the effective horizon, we can treat all coefficients as constants evaluated at \(u=u_*\) except \(\gamma^{uu}\) which vanishes at \(u=u_*\).
Then, we can partially decouple the fields by introducing new fields: \(\hat{\Phi} = \hat{P}^{\frac{1}{2}} U^{T}\tilde{\Phi}\) where \(\hat{P}^{\frac{1}{2}} \equiv \mathrm{diag}(\sqrt{p_1}, \sqrt{p_2})\).
We also define \(\hat{P}^{-\frac{1}{2}}\) as an inverse matrix of \(\hat{P}^{\frac{1}{2}}\).
In (\ref{eq:effective_action_Appendix}), relevant terms to the field behavior in the vicinity of \(u=u_*\), are only the derivative terms.
Then, the action with relevant terms is given by
\begin{equation}
	S^{(2)} = -\frac{\mathcal{N}}{2} \int \dd[4]{x}\dd{u} R
	\left[
		\gamma^{\alpha\beta}
		\partial_{\alpha}\hat{\Phi}^T \partial_{\beta} \hat{\Phi}
		+
  		2 \hat{\Phi}^T
  		\hat{P}^{-\frac{1}{2}} \hat{B}_{A}^{\alpha} \hat{P}^{-\frac{1}{2}}
  		\partial_{\alpha} \hat{\Phi}
  		+ \cdots
	\right],
	\label{eq:effective_action_relevant}
\end{equation}
where \(\hat{B}_{A}^{\alpha}\) denotes the antisymmetric part of \(U^T\tilde{B}^{\alpha}U\) which is given by
\begin{equation}
	\hat{B}_A^{\alpha} = \frac{3}{2} M^{\alpha x} \tan\theta
	\begin{bmatrix}
		0 & 1\\
		-1 & 0
	\end{bmatrix}.
\end{equation}
We ignored the symmetric part of \(U^T\tilde{B}^{\alpha}U\) and the term with \(Q\) by integration by parts.
Now, we introduce the tortoise coordinates of the effective metric:
\begin{equation}
	\tau = t - \int^u_0 \frac{\gamma^{tu}}{\gamma^{uu}} \dd u',~~~
	\sigma = \int^u_0 \frac{\sqrt{-\gamma^{tt}\gamma^{uu} + (\gamma^{tu})^2}}{\gamma^{uu}} \dd u'.
	\label{eq:tortoise_coordinates}
\end{equation}
In these coordinates, a relevant part of the first term in (\ref{eq:effective_action_relevant}) is given by
\begin{equation}
	\gamma^{\alpha\beta}\partial_{\alpha}\hat{\Phi}^T \partial_{\beta} \hat{\Phi}
	=
	\frac{-\gamma^{tt}\gamma^{uu} + (\gamma^{tu})^2}{\gamma^{uu}}
	\left[
		- \partial_{\tau} \hat{\Phi}^T \partial_{\tau} \hat{\Phi}
		+ \partial_{\sigma} \hat{\Phi}^T \partial_{\sigma} \hat{\Phi}
	\right] + \cdots
	\label{eq:decoupled_kinetic_terms}
\end{equation}
A relevant contribution from the second term in (\ref{eq:effective_action_relevant}) is given by
\begin{equation}
\begin{aligned}
	2 \hat{\Phi}^T&
	\hat{P}^{-\frac{1}{2}} \hat{B}_{A}^{\alpha} \hat{P}^{-\frac{1}{2}}
	\partial_{\alpha} \hat{\Phi}\\
	=&
	-\frac{3 M^{xu} \tan \theta}{\sqrt{p_1 p_2}}
	\hat{\Phi}^T
	\begin{bmatrix}
		0 & 1\\
		-1 & 0
	\end{bmatrix}
	\left(
		\frac{\sqrt{-\gamma^{tt}\gamma^{uu}+(\gamma^{tu})^2}}{\gamma^{uu}}
		\partial_{\sigma} \hat{\Phi}
		- \frac{\gamma^{tu}}{\gamma^{uu}}
		\partial_{\tau} \hat{\Phi}
	\right) + \cdots
\end{aligned}
\end{equation}
Notably, we can write
\(
	p_1 p_2 = \gamma^{xx} - (M^{xu})^2 \theta'(u_*)^2 = d
\), and \(d\) is defined in eq.~(\ref{eq:coeffi_characteristic}).
We obtain 
\begin{align}
	S^{(2)} =
	\frac{\mathcal{N}}{2} \int \dd[3]{x}\dd{\tau}\dd{\sigma}
	R \gamma^{tu}
	\Bigg[&
	-\partial_{\tau} \hat{\Phi}^T \partial_{\tau} \hat{\Phi}
	+\partial_{\sigma} \hat{\Phi}^T \partial_{\sigma} \hat{\Phi} \notag\\
	&+
	\frac{3 M^{xu} \tan \theta}{\gamma^{tu}\sqrt{d}}
	\hat{\Phi}^T
	\begin{bmatrix}
		0 & 1\\
		-1 & 0
	\end{bmatrix}
	\left(
		\partial_{\tau} \hat{\Phi}
		+ \partial_{\sigma} \hat{\Phi}
	\right) + \cdots
	\Bigg],
	\label{eq:eff_action_near_horizon}
\end{align}
where \(\cdots\) denotes terms of \(\order{\gamma^{uu}}\) in the vicinity of the effective horizon.%
\footnote{
	Be aware that \(\sqrt{(\gamma^{tu})^2} = - \gamma^{tu}\) because \(\gamma^{tu}\) is negative.
}

Finally, we rewrite the fields as a complex field, as follows:
\begin{equation}
	\Psi \equiv e^{i\alpha(\tau - \sigma)/2} (\hat{\Phi}_1 + i \hat{\Phi}_2)~~~
	\text{with}~~~
	\alpha \equiv
	- \frac{3 M^{xu} \tan \theta}{\gamma^{tu}\sqrt{d}}.
\end{equation}
In terms of \(\Psi\), the relevant part of the action is written as a simple form:
\begin{equation}
	S^{(2)} =
	\frac{\mathcal{N}}{2} \int \dd[3]{x}\dd{\tau}\dd{\sigma}
	R \gamma^{tu}\left[
		- | \partial_{\tau} \Psi |^2
		+ | \partial_{\sigma} \Psi |^2 + \cdots
	\right]
\end{equation}
Therefore, the equation of motion for \(\Psi\) is given by
\begin{equation}
	-\partial_{\tau}^2 \Psi + \partial_{\sigma}^2 \Psi \approx 0.
\end{equation}
We write solutions of the equation of motion and its complex conjugate, respectively, as%
\footnote{
	We can also write \(\Psi_{\pm,2}(\hat{\omega}) = \Psi_{\pm,1}^*(-\hat{\omega})\).
}
\begin{equation}
	\Psi = \int \frac{\dd{\hat{\omega}}}{2\pi} \Psi_{\pm,1}(\hat{\omega}) e^{-i\hat{\omega}(\tau \pm \sigma)},~~~
	\Psi^* = \int \frac{\dd{\hat{\omega}}}{2\pi} \Psi_{\pm,2}(\hat{\omega}) e^{-i\hat{\omega}(\tau \pm \sigma)}.
	\label{eq:solutions_complex_fields}
\end{equation}
In the vicinity of the effective horizon, the relation between the tortoise coordinates and \(t\)-\(u\) coordinates is given by
\begin{equation}
	\tau \approx t - \frac{1}{2\kappa} \log\left|1 - \frac{u}{u_*}\right|,~~~
	{\sigma}\approx  - \frac{1}{2\kappa} \log\left|1 - \frac{u}{u_*}\right|,
\end{equation}
where we have used eq.~(\ref{eq:surface_gravity_in_gamma-Inv}) for rewriting $\kappa$ which is the surface gravity on the effective horizon $u=u_*$ defined by eq.~(\ref{eq:surface_gravity}).
Because \(u=u_*\) corresponds to \(\sigma = \infty\), a choice of the positive/negative sign in (\ref{eq:solutions_complex_fields}) corresponds to the outgoing/ingoing-wave solution at the effective horizon.
In terms of \(\Psi_{\pm,s}\), the real field \(\hat{\Phi}_1\) is written as
\begin{equation}
\begin{aligned}
	\hat{\Phi}_1
	=&
	\frac{1}{2}\left(
		e^{-i\tfrac{\alpha}{2}(\tau - \sigma)} \Psi
		+
		e^{i\tfrac{\alpha}{2}(\tau - \sigma)} \Psi^*
	\right)\\
	\approx&
	\int \frac{\dd{\omega}}{2\pi} e^{-i\omega t} \frac{1}{2}\left[
		\Psi_{\pm,1} \left(1-\frac{u}{u_*}\right)^{i\lambda_{\pm,1}}
		+ \Psi_{\pm,2} \left(1-\frac{u}{u_*}\right)^{i\lambda_{\pm,2}}
	\right],
	\label{eq:solution_Phi1_near_horizon}
\end{aligned}
\end{equation}
in the vicinity of \(u=u_*\) where
\begin{equation}
\begin{aligned}
	\lambda_{\pm,s}
	=&
	\left( \frac{1\pm1}{2} \right)
	\frac{2\omega + (-1)^s \alpha}{\kappa}\\
	=&
	\frac{1\pm1}{2}
	\left(
		\frac{\omega}{\kappa}
		-(-1)^s \frac{3M^{xu}\tan\theta(u_*)/\sqrt{d}}{(\gamma^{uu})'(u_*)}
	\right),
\end{aligned}
\label{eq:lambda_pm_s}
\end{equation}
for \(s = 1,2\).
We can also obtain a similar result for \(\hat{\Phi}_2\).
The original fields \(\Phi^I\) are given by linear combinations of \(\hat{\Phi}^I\).
The four exponents of (\ref{eq:lambda_pm_s}) completely match to those of (\ref{eq:Frobenius_exponents}).
Consequently, we can understand that \(\lambda_{-,s} = 0\)
for $s=1,2$ corresponds to the set of the ingoing-wave solutions manifestly
because the negative sign has been taken in eq.~(\ref{eq:solutions_complex_fields}).

\bibliography{main}

\providecommand{\href}[2]{#2}\begingroup\raggedright\begin{thebibliography}{10}

\bibitem{Maldacena:1998}
J.M.~Maldacena, \emph{The {large N} limit of superconformal field theories and
  supergravity}, {\emph{Adv. Theor. Math. Phys.} {\bfseries 2} (1998) 231}
  [\href{https://arxiv.org/abs/hep-th/9711200}{{\ttfamily hep-th/9711200}}].

\bibitem{Gubser:1998}
S.S.~Gubser, I.R.~Klebanov and A.M.~Polyakov, \emph{Gauge theory correlators
  from non-critical string theory}, {\emph{Phys. Lett. B} {\bfseries 428}
  (1998) } [\href{https://arxiv.org/abs/hep-th/9802109}{{\ttfamily
  hep-th/9802109}}].

\bibitem{Witten:1998}
E.~Witten, \emph{Anti-de sitter space and holography}, {\emph{Adv. Theor. Math.
  Phys.} {\bfseries 2} (1998) }
  [\href{https://arxiv.org/abs/hep-th/9802150}{{\ttfamily hep-th/9802150}}].

\bibitem{Liu:2018crr}
H.~Liu and J.~Sonner, \emph{{Holographic systems far from equilibrium: a
  review}},  \href{https://arxiv.org/abs/1810.02367}{{\ttfamily 1810.02367}}.

\bibitem{Karch:2002sh}
A.~Karch and E.~Katz, \emph{{Adding flavor to AdS / CFT}},
  \href{https://doi.org/10.1088/1126-6708/2002/06/043}{\emph{JHEP} {\bfseries
  06} (2002) 043} [\href{https://arxiv.org/abs/hep-th/0205236}{{\ttfamily
  hep-th/0205236}}].

\bibitem{Karch:2007pd}
A.~Karch and A.~O'Bannon, \emph{{Metallic AdS/CFT}},
  \href{https://doi.org/10.1088/1126-6708/2007/09/024}{\emph{JHEP} {\bfseries
  09} (2007) 024} [\href{https://arxiv.org/abs/0705.3870}{{\ttfamily
  0705.3870}}].

\bibitem{Erdmenger:2007bn}
J.~Erdmenger, R.~Meyer and J.P.~Shock, \emph{{AdS/CFT with flavour in electric
  and magnetic Kalb-Ramond fields}},
  \href{https://doi.org/10.1088/1126-6708/2007/12/091}{\emph{JHEP} {\bfseries
  12} (2007) 091} [\href{https://arxiv.org/abs/0709.1551}{{\ttfamily
  0709.1551}}].

\bibitem{Albash:2007bq}
T.~Albash, V.G.~Filev, C.V.~Johnson and A.~Kundu, \emph{{Quarks in an external
  electric field in finite temperature large N gauge theory}},
  \href{https://doi.org/10.1088/1126-6708/2008/08/092}{\emph{JHEP} {\bfseries
  08} (2008) 092} [\href{https://arxiv.org/abs/0709.1554}{{\ttfamily
  0709.1554}}].

\bibitem{Bergman:2008sg}
O.~Bergman, G.~Lifschytz and M.~Lippert, \emph{{Response of Holographic QCD to
  Electric and Magnetic Fields}},
  \href{https://doi.org/10.1088/1126-6708/2008/05/007}{\emph{JHEP} {\bfseries
  05} (2008) 007} [\href{https://arxiv.org/abs/0802.3720}{{\ttfamily
  0802.3720}}].

\bibitem{Hashimoto:2013mua}
K.~Hashimoto and T.~Oka, \emph{{Vacuum Instability in Electric Fields via
  AdS/CFT: Euler-Heisenberg Lagrangian and Planckian Thermalization}},
  \href{https://doi.org/10.1007/JHEP10(2013)116}{\emph{JHEP} {\bfseries 10}
  (2013) 116} [\href{https://arxiv.org/abs/1307.7423}{{\ttfamily 1307.7423}}].

\bibitem{Nakamura:2010zd}
S.~Nakamura, \emph{{Negative Differential Resistivity from Holography}},
  \href{https://doi.org/10.1143/PTP.124.1105}{\emph{Prog. Theor. Phys.}
  {\bfseries 124} (2010) 1105}
  [\href{https://arxiv.org/abs/1006.4105}{{\ttfamily 1006.4105}}].

\bibitem{Mateos:2007vn}
D.~Mateos, R.C.~Myers and R.M.~Thomson, \emph{{Thermodynamics of the brane}},
  \href{https://doi.org/10.1088/1126-6708/2007/05/067}{\emph{JHEP} {\bfseries
  05} (2007) 067} [\href{https://arxiv.org/abs/hep-th/0701132}{{\ttfamily
  hep-th/0701132}}].

\bibitem{Kaminski:2009ce}
M.~Kaminski, K.~Landsteiner, F.~Pena-Benitez, J.~Erdmenger, C.~Greubel and
  P.~Kerner, \emph{{Quasinormal modes of massive charged flavor branes}},
  \href{https://doi.org/10.1007/JHEP03(2010)117}{\emph{JHEP} {\bfseries 03}
  (2010) 117} [\href{https://arxiv.org/abs/0911.3544}{{\ttfamily 0911.3544}}].

\bibitem{Horowitz:1999jd}
G.T.~Horowitz and V.E.~Hubeny, \emph{{Quasinormal modes of AdS black holes and
  the approach to thermal equilibrium}},
  \href{https://doi.org/10.1103/PhysRevD.62.024027}{\emph{Phys. Rev. D}
  {\bfseries 62} (2000) 024027}
  [\href{https://arxiv.org/abs/hep-th/9909056}{{\ttfamily hep-th/9909056}}].

\bibitem{Nunez:2003eq}
A.~Nunez and A.O.~Starinets, \emph{{AdS / CFT correspondence, quasinormal
  modes, and thermal correlators in N=4 SYM}},
  \href{https://doi.org/10.1103/PhysRevD.67.124013}{\emph{Phys. Rev. D}
  {\bfseries 67} (2003) 124013}
  [\href{https://arxiv.org/abs/hep-th/0302026}{{\ttfamily hep-th/0302026}}].

\bibitem{Kovtun:2005ev}
P.K.~Kovtun and A.O.~Starinets, \emph{{Quasinormal modes and holography}},
  \href{https://doi.org/10.1103/PhysRevD.72.086009}{\emph{Phys. Rev. D}
  {\bfseries 72} (2005) 086009}
  [\href{https://arxiv.org/abs/hep-th/0506184}{{\ttfamily hep-th/0506184}}].

\bibitem{Kaminski:2009dh}
M.~Kaminski, K.~Landsteiner, J.~Mas, J.P.~Shock and J.~Tarrio,
  \emph{{Holographic Operator Mixing and Quasinormal Modes on the Brane}},
  \href{https://doi.org/10.1007/JHEP02(2010)021}{\emph{JHEP} {\bfseries 02}
  (2010) 021} [\href{https://arxiv.org/abs/0911.3610}{{\ttfamily 0911.3610}}].

\bibitem{Schoell:2001}
E.~Sch\"oll, \emph{Nonlinear Spatio-Temporal Dynamics and Chaos in
  Semiconductors}, Cambridge University Press (2001).

\bibitem{Hashimoto:2014yza}
K.~Hashimoto, S.~Kinoshita, K.~Murata and T.~Oka, \emph{{Electric Field Quench
  in AdS/CFT}}, \href{https://doi.org/10.1007/JHEP09(2014)126}{\emph{JHEP}
  {\bfseries 09} (2014) 126} [\href{https://arxiv.org/abs/1407.0798}{{\ttfamily
  1407.0798}}].

\bibitem{Kim:2011qh}
K.-Y.~Kim, J.P.~Shock and J.~Tarrio, \emph{{The open string membrane paradigm
  with external electromagnetic fields}},
  \href{https://doi.org/10.1007/JHEP06(2011)017}{\emph{JHEP} {\bfseries 06}
  (2011) 017} [\href{https://arxiv.org/abs/1103.4581}{{\ttfamily 1103.4581}}].

\bibitem{Seiberg:1999vs}
N.~Seiberg and E.~Witten, \emph{{String theory and noncommutative geometry}},
  \href{https://doi.org/10.1088/1126-6708/1999/09/032}{\emph{JHEP} {\bfseries
  09} (1999) 032} [\href{https://arxiv.org/abs/hep-th/9908142}{{\ttfamily
  hep-th/9908142}}].

\bibitem{Mateos:2006nu}
D.~Mateos, R.C.~Myers and R.M.~Thomson, \emph{{Holographic phase transitions
  with fundamental matter}},
  \href{https://doi.org/10.1103/PhysRevLett.97.091601}{\emph{Phys. Rev. Lett.}
  {\bfseries 97} (2006) 091601}
  [\href{https://arxiv.org/abs/hep-th/0605046}{{\ttfamily hep-th/0605046}}].

\bibitem{Frolov:2006tc}
V.P.~Frolov, \emph{{Merger Transitions in Brane-Black-Hole Systems:
  Criticality, Scaling, and Self-Similarity}},
  \href{https://doi.org/10.1103/PhysRevD.74.044006}{\emph{Phys. Rev. D}
  {\bfseries 74} (2006) 044006}
  [\href{https://arxiv.org/abs/gr-qc/0604114}{{\ttfamily gr-qc/0604114}}].

\bibitem{Kruczenski:2003be}
M.~Kruczenski, D.~Mateos, R.C.~Myers and D.J.~Winters, \emph{{Meson
  spectroscopy in AdS / CFT with flavor}},
  \href{https://doi.org/10.1088/1126-6708/2003/07/049}{\emph{JHEP} {\bfseries
  07} (2003) 049} [\href{https://arxiv.org/abs/hep-th/0304032}{{\ttfamily
  hep-th/0304032}}].

\bibitem{Kirsch:2006he}
I.~Kirsch, \emph{{Spectroscopy of fermionic operators in AdS/CFT}},
  \href{https://doi.org/10.1088/1126-6708/2006/09/052}{\emph{JHEP} {\bfseries
  09} (2006) 052} [\href{https://arxiv.org/abs/hep-th/0607205}{{\ttfamily
  hep-th/0607205}}].

\bibitem{Abt:2019tas}
R.~Abt, J.~Erdmenger, N.~Evans and K.S.~Rigatos, \emph{{Light composite
  fermions from holography}},
  \href{https://doi.org/10.1007/JHEP11(2019)160}{\emph{JHEP} {\bfseries 11}
  (2019) 160} [\href{https://arxiv.org/abs/1907.09489}{{\ttfamily
  1907.09489}}].

\bibitem{Hashimoto:2015wpa}
K.~Hashimoto, S.~Kinoshita and K.~Murata, \emph{{Conic D-branes}},
  \href{https://doi.org/10.1093/ptep/ptv105}{\emph{PTEP} {\bfseries 2015}
  (2015) 083B04} [\href{https://arxiv.org/abs/1505.04506}{{\ttfamily
  1505.04506}}].

\bibitem{Mas:2008jz}
J.~Mas, J.P.~Shock, J.~Tarrio and D.~Zoakos, \emph{{Holographic Spectral
  Functions at Finite Baryon Density}},
  \href{https://doi.org/10.1088/1126-6708/2008/09/009}{\emph{JHEP} {\bfseries
  09} (2008) 009} [\href{https://arxiv.org/abs/0805.2601}{{\ttfamily
  0805.2601}}].

\bibitem{Mas:2009wf}
J.~Mas, J.P.~Shock and J.~Tarrio, \emph{{Holographic Spectral Functions in
  Metallic AdS/CFT}},
  \href{https://doi.org/10.1088/1126-6708/2009/09/032}{\emph{JHEP} {\bfseries
  09} (2009) 032} [\href{https://arxiv.org/abs/0904.3905}{{\ttfamily
  0904.3905}}].

\bibitem{Ishigaki:2020coe}
S.~Ishigaki and S.~Nakamura, \emph{{Mechanism for negative differential
  conductivity in holographic conductors}},
  \href{https://doi.org/10.1007/JHEP12(2020)124}{\emph{JHEP} {\bfseries 12}
  (2020) 124} [\href{https://arxiv.org/abs/2008.00904}{{\ttfamily
  2008.00904}}].

\bibitem{Ishigaki:2020vtr}
S.~Ishigaki and M.~Matsumoto, \emph{{Nambu-Goldstone modes in non-equilibrium
  systems from AdS/CFT correspondence}},
  \href{https://doi.org/10.1007/JHEP04(2021)040}{\emph{JHEP} {\bfseries 04}
  (2021) 040} [\href{https://arxiv.org/abs/2012.01177}{{\ttfamily
  2012.01177}}].

\bibitem{Amado:2009ts}
I.~Amado, M.~Kaminski and K.~Landsteiner, \emph{{Hydrodynamics of Holographic
  Superconductors}},
  \href{https://doi.org/10.1088/1126-6708/2009/05/021}{\emph{JHEP} {\bfseries
  05} (2009) 021} [\href{https://arxiv.org/abs/0903.2209}{{\ttfamily
  0903.2209}}].

\bibitem{Karch:2005ms}
A.~Karch, A.~O'Bannon and K.~Skenderis, \emph{{Holographic renormalization of
  probe D-branes in AdS/CFT}},
  \href{https://doi.org/10.1088/1126-6708/2006/04/015}{\emph{JHEP} {\bfseries
  04} (2006) 015} [\href{https://arxiv.org/abs/hep-th/0512125}{{\ttfamily
  hep-th/0512125}}].

\bibitem{Sornette_1998}
D.~Sornette, \emph{Discrete-scale invariance and complex dimensions},
  \href{https://doi.org/10.1016/s0370-1573(97)00076-8}{\emph{Physics Reports}
  {\bfseries 297} (1998) 239–270}.

\bibitem{Matsumoto:2018ukk}
M.~Matsumoto and S.~Nakamura, \emph{{Critical Exponents of Nonequilibrium Phase
  Transitions in AdS/CFT Correspondence}},
  \href{https://doi.org/10.1103/PhysRevD.98.106027}{\emph{Phys. Rev. D}
  {\bfseries 98} (2018) 106027}
  [\href{https://arxiv.org/abs/1804.10124}{{\ttfamily 1804.10124}}].

\bibitem{Imaizumi:2019byu}
T.~Imaizumi, M.~Matsumoto and S.~Nakamura, \emph{{Current Driven Tricritical
  Point in Large- $N_c$ Gauge Theory}},
  \href{https://doi.org/10.1103/PhysRevLett.124.191603}{\emph{Phys. Rev. Lett.}
  {\bfseries 124} (2020) 191603}
  [\href{https://arxiv.org/abs/1911.06262}{{\ttfamily 1911.06262}}].

\bibitem{Banerjee:2015cvy}
A.~Banerjee, A.~Kundu and S.~Kundu, \emph{{Flavour Fields in Steady State:
  Stress Tensor and Free Energy}},
  \href{https://doi.org/10.1007/JHEP02(2016)102}{\emph{JHEP} {\bfseries 02}
  (2016) 102} [\href{https://arxiv.org/abs/1512.05472}{{\ttfamily
  1512.05472}}].

\bibitem{Nakamura:2009tf}
S.~Nakamura, H.~Ooguri and C.-S.~Park, \emph{{Gravity Dual of Spatially
  Modulated Phase}},
  \href{https://doi.org/10.1103/PhysRevD.81.044018}{\emph{Phys. Rev. D}
  {\bfseries 81} (2010) 044018}
  [\href{https://arxiv.org/abs/0911.0679}{{\ttfamily 0911.0679}}].

\bibitem{Bergman:2011rf}
O.~Bergman, N.~Jokela, G.~Lifschytz and M.~Lippert, \emph{{Striped instability
  of a holographic Fermi-like liquid}},
  \href{https://doi.org/10.1007/JHEP10(2011)034}{\emph{JHEP} {\bfseries 10}
  (2011) 034} [\href{https://arxiv.org/abs/1106.3883}{{\ttfamily 1106.3883}}].

\end{thebibliography}\endgroup
\bibliographystyle{JHEP}
\end{document}